\documentclass[12pt]{article}
\usepackage{setspace}
\usepackage{indentfirst}
\usepackage{natbib}
\usepackage{xltabular}
\usepackage{booktabs}
\usepackage{array}
\usepackage{subcaption}
\newcolumntype{L}[1]{>{\raggedright\arraybackslash}p{#1}}
\newcolumntype{R}[1]{>{\raggedleft\arraybackslash}p{#1}}
\usepackage{amsmath}
\usepackage{ragged2e}
\usepackage{authblk}
\usepackage{amsmath}
\usepackage{breqn}
\usepackage{graphics}
\usepackage{amsfonts}
\usepackage{graphicx}
\usepackage{amsmath}
\usepackage{wrapfig}
\usepackage{booktabs,caption}
\usepackage{adjustbox}
\usepackage{enumitem}
\usepackage[flushleft]{threeparttable}
\usepackage[colorlinks,
            linkcolor=blue,
            anchorcolor=blue,
            citecolor=blue]{hyperref}
            
\usepackage[left=1.in, right=1.in, top=.5in, bottom=.5in, includehead, includefoot]{geometry}       

\usepackage{floatrow}
\usepackage[multiple]{footmisc}
\usepackage{caption}
\begin{document}

\begin{titlepage}
\title{The Effects of Innovation on Foreign Portfolio Investment: The Role of Institutions and Risk-Taking
\thanks{We appreciate valuable comments by Xiangyu Feng, Yinong Tan, Yong Wang, Qi Xu, and Zilong Zhang. This study was supported by the Ishii Memorial Securities Research Promotion Foundation (Grant-in-Aid for Research, 2025). Corresponding author: Tomoo Kikuchi. Nishi-Waseda Bldg.7F, 1-21-1 Nishi-Waseda, Shinjyuku-ku, Tokyo 169-0051 Japan. Email: \href{mailto:tomookikuchi@waseda.jp}{tomookikuchi@waseda.jp}. }}
\author[a]{Yimin Wu}
\author[b]{Tomoo Kikuchi}
\affil[a,b]{Graduate School of Asia-Pacific Studies, Waseda University}
\date{\today}
\maketitle

\vspace{-1.5cm}
\begin{abstract}
\begin{spacing}{1}
    \noindent We study whether and how innovation intensity attracts foreign portfolio investment (FPI) using a panel of 60 countries from 1996 to 2021. Using an instrumental variable strategy based on regional shift-share and global push instruments, we estimate the causal response of debt and equity inflows to innovation intensity in the host country. We find that innovation increases FPI, with larger effects for equity than debt inflow. Moreover, the effect of innovation on equity inflow increases with technological development and institutional quality, whereas the effect on debt inflow is positive and significant only at high levels of these factors. We also find that countries with a higher risk-taking environment attract more FPI and that equity inflow responses are immediate and persistent, whereas debt inflow responses are modest and dampen over time.

\end{spacing}
\vspace{0.5cm}
\noindent\textbf{Keywords:} Innovation intensity; foreign portfolio investment; institutions; technological development

\end{abstract}

\setcounter{page}{0}
\thispagestyle{empty}

\end{titlepage}

\pagebreak \newpage

\section{Introduction} \label{sec:introduction}

Foreign portfolio investment (FPI), defined as the net purchase of equities and debt securities by foreign investors, has become a major component of international capital flows over the past decades by allowing international risk sharing and by providing recipient countries with access to a broad investor base through equity and debt markets. 
This paper finds that innovation intensity is a key determinant of FPI. The central mechanism is that innovation raises expected future returns and growth opportunities by capturing forward-looking information about productivity growth, firm valuation, and investable assets. Thus, countries with stronger innovation capacity attract more FPI, especially in equity whose returns are closely tied to expectations about future cash flows and technological progress.\footnote{Innovation relies more on the equity market than the debt market for finance \citep{raghuram1998dependence,allen1999diversity,brown2013law}. This literature motivates the equity-oriented financing channel, while our paper studies the reverse direction from innovation to foreign portfolio allocation.}

\begin{figure}[ht!] 
	\begin{center} 
		\includegraphics[width=.8\textwidth]{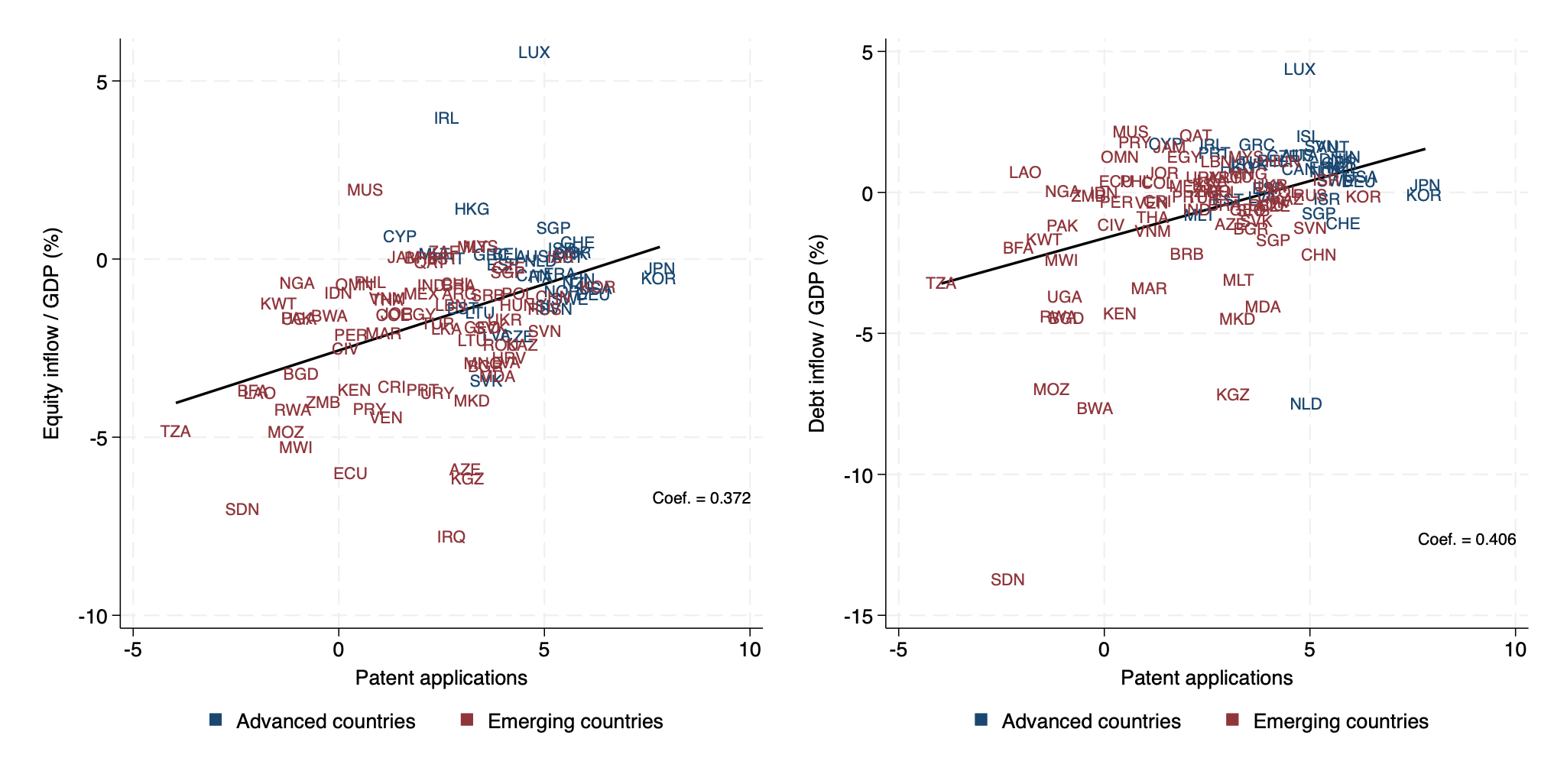} 
		\caption{The Number of Patent Applications (Per Million People) vs Portfolio Inflows} 
		\label{figure:scatter} \end{center} 
	\floatfoot{
		\footnotesize\textit{Note}: Each plot is the annual 1980-2021 average for each country. There are 24 advanced economies and 73 emerging economies.}		
\end{figure}

Figure \ref{figure:scatter}  shows scatter plots of the number of patent applications and FPI (both debt and equity inflows). Two features stand out: first, the number of patent applications is indeed positively associated with FPI;  second, advanced economies have both higher FPI and number of patent applications than emerging economies.
These observations suggest that the positive relationship between innovation and FPI depends on additional moderating conditions. 

We show that the effect of innovation on FPI depends critically on institutional quality and differs for debt and equity inflows. Debt inflow is typically associated with longer investment horizons, lower turnover, and a greater emphasis on capital preservation. Hence, debt inflow is highly sensitive to institutional risks related to property rights, legal enforcement, and sovereign credibility, and does not change destinations frequently in response to short-term signals \citep{burger2007foreign}. Innovation, therefore, attracts debt inflow only in environments where institutional quality is sufficiently high to support long-term commitments.

In contrast, equity inflow is more flexible, liquid, and responsive to changes in expected returns. Equity investors can quickly rebalance portfolios and are more willing to find growth opportunities even in environments with institutional frictions. This behavior is consistent with evidence that equity flows are more volatile, more return-seeking, and more sensitive to news about growth prospects and valuation \citep{froot2001portfolio}. Therefore, innovation activity can attract equity inflow at a much lower level of institutional quality than that required to attract debt inflow.

We use a panel of up to 159 countries from 1996 to 2021 (reduced to 60 countries after including a set of controls). Equity and debt inflows are measured as a percentage of GDP. Innovation intensity is measured by the number of patent applications per million people and instrumented using a combination of push-type instruments. The global push instrument measures a country’s research capacity relative to the world frontier, proxied by the ratio of domestic researchers to the maximum number of researchers observed in the same year. We also construct a regional innovation spillover instrument from regional patent growth interacted with the geographical distance to the regional innovation leader. Estimation is conducted using  two-stage least squares (2SLS) with region and time fixed effects and a set of macro-financial controls. First-stage results indicate strong instrument relevance.

Baseline results show that innovation intensity positively affects both equity and debt inflows. However, once institutional or technological  interactions are introduced, strong composition effects emerge; innovation intensity significantly increases equity inflow across a wide range of institutional or technological environments, while its impact on debt inflow is significant only in countries with very high institutional quality or technological development. 
Furthermore, local projection analyses show that innovation shocks have persistent effects on portfolio inflows, particularly for equity inflow, consistent with forward-looking investor behavior.

Our paper contributes to different strands of literature. First, there is a large body of literature starting with \cite{lucas1990} investigating why capital does not flow to high-return destinations. The literature includes works that emphasize domestic fundamentals such as  limited complementary inputs, weak institutions, underdeveloped financial markets, or capital controls \citep{bekaert2005does, Prasad2007,alfaro2008doesn, bekaert2000foreign,henry2000stock}, international frictions such as sovereign risk, expropriation risk, and information asymmetries \citep{brennan1997international,portes2005determinants}, or  
global financial conditions, especially those linked to US monetary policy, which drive synchronized surges and reversals in capital flows across countries \citep{forbes2012capital,rey2015dilemma}. 

This paper contributes to the literature by introducing innovation intensity as a forward-looking pull factor that shapes both the level and composition of FPI. In doing so, we provide a complementary explanation for the portfolio investment dimension of the Lucas paradox by showing that global investors allocate capital toward countries with stronger innovation intensity. We find that innovation-intensive economies, which are mostly advanced economies, attract significantly more FPI than less innovation-intensive ones. This is consistent with the fact that many emerging economies export capital even when they grow faster than advanced economies \citep{gourinchas2013capital}.

Second, there is a strand in the literature that focuses on heterogeneous factors attracting different asset classes; equity inflow is more sensitive to growth opportunities and investor protection, whereas debt inflow depends more on currency denomination, bond market development, and investor base characteristics \citep{eichengreen1999exchange,burger2007foreign,ma2020international}. These patterns suggest that long-run portfolio allocation depends on forward-looking assessments of risk and transparency.
We contribute to the literature by highlighting a clear asymmetry between equity and debt flows;  innovation has a strong and robust effect on equity inflow, but a much weaker effect on debt inflow. Moreover, the equity-to-debt (E/D) inflow ratio increases as innovation activity rises, indicating that foreign investors primarily respond to innovation through equity markets.

Third, \cite{alfaro2008doesn} and the subsequent works highlight the role of institutions, showing that poor governance, corruption, weak legal enforcement, and insecure property rights all raise uncertainty and hinder FPI.  They show that once institutional quality is accounted for, income per capita explains far less of the variation in capital flows, suggesting that institutions are a key driver of the Lucas paradox. Strong institutions reduce risk and improve contract enforcement, thereby increasing the risk-adjusted attractiveness of domestic assets to foreign investors \citep{porta1998law,alfaro2008doesn,leuz2009foreigners}, whereas capital inflows may in turn enhance institutional development by promoting financial deepening and market discipline \citep{bekaert2005does, igan2022capital}. We uncover a nonlinear interaction between innovation and institutional quality. We show that there is a development-ladder process for innovation to attract equity inflow; the effect of innovation on equity inflow becomes larger and more significant as institutional quality improves, whereas there is a threshold effect for innovation to attract debt inflow; the effect on debt inflow is significant only in countries with an extremely high level of institutional quality. 
These nonlinearities are less emphasized in existing studies, which typically assume linear or symmetric effects of institutions on capital flows.

Finally, there are papers that study whether foreign portfolio equity participation promotes innovation at the firm level \citep{Bena2017, Luong2017, Yi2023} and the industry level \citep{Moshirian2021, Wang2022}. 
%
Overall, this literature suggests that foreign portfolio equity participation can foster innovation by easing financing constraints, improving governance, strengthening risk sharing, and transmitting knowledge and information. Our paper is distinct in that we study the reserve causality showing that innovation leads to both equity and debt inflows, and that this relationship is moderated by technological development and institutional quality.\footnote{\cite{kikuchi2018volatile} develops a theoretical model where entrepreneurial activity expands attracting foreign capital flows in a boom phase and contracts leading to capital flow reversals in a bust phase. The relationship between entrepreneurial activity and foreign capital inflow is moderated by institutional quality (degree of moral hazard) and the degree of risk aversion of investors as predicted in this paper.}


The rest of the paper is organized as follows. Section \ref{data_fpi} presents the data and variable construction. Section \ref{iv_fpi} outlines the empirical strategy and baseline results. Section \ref{sec:hete} examines asymmetries regarding how portfolio investment depends on institutional quality. 
Section \ref{sec:further_fpi} presents the role of risk-taking. Section \ref{sec:dynamic_fpi} shows dynamic effects. Section \ref{sec:conclusion_fpi} concludes.

\section{Data}\label{data_fpi}

We examine the impact of innovation on FPI using an unbalanced panel of 159 advanced and emerging countries, with annual data from 1960 to 2021. The availability of the necessary instruments and control variables reduces the sample to 60 countries spanning from 1996 to 2021. We measure the innovation of each country by the number of patent applications per million people. The dependent variables are equity inflow, debt inflow, the E/D inflow ratio, and total FPI. We describe the data and their sources in Table \ref{table:data_description}.\footnote{To robustly check the baseline findings, we further estimate the results using total charges for the authorized use of intellectual property (IP) rights per million people in Section \ref{ipcharges}.} 

Table \ref{table:sum_fpi} reports the summary statistics for the full sample and for the subsamples of advanced and emerging economies. Several features stand out from the full sample. First, patenting activity is highly uneven across economies. The mean number of patent applications (per million people) is 213.16, with a large standard deviation of 458.75, and the distribution ranges from 0.20 to 3278.94. This suggests substantial cross-country heterogeneity in innovative activity. Second, foreign portfolio inflows also vary widely. Equity inflow averages 5.18 percent of GDP, while debt inflow averages 2.65 percent of GDP. At the same time, both variables display very large dispersion, with equity inflow ranging from -224.23 to 731.10 and debt inflow ranging from -33.06 to 487.65. The E/D inflow ratio has a mean of -187.46 and a standard deviation of 12858.38, again indicating considerable heterogeneity in the composition of external portfolio inflows across economies.

\begin{spacing}{0.5}
\begin{table}[ht!]
\centering
\footnotesize
\renewcommand{\arraystretch}{1} 
\begin{threeparttable}
\begin{tabular}{p{4cm} p{7cm} p{4cm}} 
\toprule
Notation & Description & Data Source \\
\midrule
\textbf{Dependent variables} & & \\
\hspace{0.2cm}$y^{s}_{i,t}$ & Equity inflow as a percentage of GDP: Portfolio investment, net incurrence of liabilities, equity securities. &   Coordinated Portfolio \newline Investment Survey (CPIS) from IMF database\\ \hspace{0.2cm}
& Debt inflow as a percentage of GDP: Portfolio investment, net incurrence of liabilities, debt securities. &  \\

& The E/D inflow ratio: Equity inflow divided by debt inflow. &\\
& The total FPI as a percentage of GDP: The sum of equity inflow and debt inflow, both measured as percentages of GDP.  &\\

\midrule
\textbf{Independent variables} & & \\
\hspace{0.2cm}$x_{i,t}$ & The number of patent applications filed through the Patent Cooperation Treaty or national offices per million people. & World Bank Database \\
\hspace{0.2cm}$TFP_{i,t}$ & Total factor productivity level measured at current PPPs (USA=1). & Penn World Tables 10.01 \\
\hspace{0.2cm}$q_{i,t}$ & One of six institutional quality measures: regulatory quality; voice and accountability; control of corruption; rule of law; political stability; or government effectiveness.& \cite{kaufmann2024worldwide} \\
\midrule
\textbf{Instruments} & & \\
\hspace{0.2cm}$z_{i,t}$ & The interaction term between the logarithm of the geographical distance between country $i$ and regional technological leader and average regional patent growth excluding country $i$. & GeoDist Database \newline World Bank Database \\
\hspace{0.2cm}$D_{i,r}$ & The set of regional dummies that equals 1 if the country $i$ belongs to region $r$, and 0 otherwise. Regions include Asia, Africa, Europe, and the Western Hemisphere. & \cite{liu2023capital} \\
\hspace{0.2cm}$n_{i,t}$ & The number of researchers working in R\&D sector in country $i$. & World Bank Database \\
\midrule
\textbf{Control variables} & & \\
\hspace{0.2cm}$Findev_{i,t}$ & Stock market capitalization as percentage of GDP. & World Bank Database \\
\hspace{0.2cm}$Tradopen_{i,t}$ & Sum of exports and imports of goods and services measured as a share of GDP. & OECD National Accounts \& World Bank Database \\
\hspace{0.2cm}$HCI_{i,t}$ & Years of schooling and returns to education. & Penn World Tables 10.01 \\
\hspace{0.2cm}$Goversize_{i,t}$ & Government final consumption expenditure excluding capital formation in defense and security. & World Bank Database \\
\hspace{0.2cm}$Inflat_{i,t}$ & Inflation measured by annual percentage change in consumer price index (Laspeyres formula). & World Bank Database \\
\hspace{0.2cm}$Gdpgrow_{i,t}$ & Real GDP growth rate. & IMF World Economic \newline Outlook Database \\
\hspace{0.2cm}$Bankcri_{i,t}$ & Dummy indicating banking crisis: 1 if crisis, 0 otherwise. Defined by significant financial distress and policy intervention. & \cite{laeven2018systemic} \\
\bottomrule
\end{tabular}
\caption{Definition and Notation of Variables}
\label{table:data_description}
\end{threeparttable}
\end{table}
\end{spacing}
\vspace{0.5cm}
\clearpage

The subsample statistics reveal clear differences between advanced and emerging economies. Advanced economies are substantially more innovative, with an average of 343.08 patents compared with only 33.75 in emerging economies. They also receive much larger foreign portfolio inflows. In advanced economies, mean equity and debt inflows amount to 8.61 percent and 3.98 percent of GDP, respectively, whereas the corresponding values for emerging economies are 0.82 percent and 0.44 percent. The E/D inflow ratio is also positive (228.64) in advanced economies but strongly negative (-800.05) in emerging economies. These patterns are consistent with the view that more innovative economies tend to be associated with larger external portfolio inflows, particularly on the equity side.

\begin{table}[ht]
        \centering
        \footnotesize
            \begin{threeparttable}
                \begin{tabular}{p{5cm} R{1.5cm} R{1.5cm} R{1.5cm} R{1.9cm} R{1.7cm}} 
                    \toprule
                    Variable & N & Mean & S.D. & Min & Max \\ [0.5ex] 
                    \midrule
                
                     \textbf{Full sample} &  &  &  &  &  \\
                    \hspace{0.2cm}Patents & 1319 & 213.155  & 458.747 & 0.199 & 3278.941 \\
                     \hspace{0.2cm}Equity inflow & 1308 & 5.180  & 47.008 & -224.232 & 731.100 \\
                    \hspace{0.2cm}Debt inflow & 1319 & 2.652 & 16.500 & -33.062 & 487.645 \\
                    \hspace{0.2cm}E/D inflow ratio & 1246 & -187.459 & 12858.380 & -404150.000 & 189874.600 \\
                     \midrule
                    \textbf{Sub sample} &  &  &  &  &  \\
                    \hspace{0.2cm}Advanced economies &  &  &  &  &  \\
                    \hspace{0.5cm}Patents & 765 & 343.079  & 564.197 & 0.829 & 3278.941 \\
                    \hspace{0.5cm}Equity inflow & 759 & 8.611  & 61.447 & -224.232 & 731.100 \\
                    \hspace{0.5cm}Debt inflow & 765 & 3.982 & 21.446 & -33.062 & 487.645 \\
                     \hspace{0.5cm}E/D inflow ratio & 742 & 228.642 & 7574.498 & -52568.250 & 189874.600 \\
                    \hspace{0.2cm}Emerging economies &  &  &  &  &  \\
                    \hspace{0.5cm}Patents & 554 & 33.748  & 78.778 & 0.199 & 892.204 \\
                    \hspace{0.5cm}Equity inflow& 549 & 0.435  & 2.976 & -30.026 & 51.357 \\
                    \hspace{0.5cm}Debt inflow & 554 & 0.817 & 2.768 & -30.461 & 25.464 \\
                     \hspace{0.5cm}E/D inflow ratio & 504 & -800.053 & 18002.360 & -404150.000 & 209.012 \\
                    \bottomrule
                \end{tabular}
                \begin{tablenotes}
                    \small
                    \item Note: Summary statistics of the data sample for the baseline regressions.
                \end{tablenotes}
                \caption{Summary Statistics}
                \label{table:sum_fpi}
            \end{threeparttable}
    \end{table}

Finally, the wide minima and maxima in both the full sample and the subsamples point to substantial cross-country and intertemporal variation in external portfolio structure. This is particularly evident for debt and equity inflows, whose extreme values indicate that some economies experienced unusually large inflows relative to GDP. The contrast between advanced and emerging economies is also evident in the dispersion of patenting activity and portfolio inflows, with advanced economies exhibiting both higher means and greater variation. Overall, the summary statistics provide descriptive support for the main hypothesis of the paper: economies with stronger innovative activity, particularly advanced economies, tend to have larger foreign portfolio inflows and a composition that is relatively more tilted toward equity.

\section{The Panel Instrumental Variable Approach}\label{iv_fpi}

This section reports our main results on how innovation attracts FPI. We regress the log of equity inflow and debt inflow  (percent of GDP) on innovation intensity, proxied by the logged patent applications per million people, while controlling for financial development, trade openness, human capital, government size, inflation, real growth, and banking crises (Table \ref{table:data_description}). We employ 2SLS to address endogeneity, especially the possibility that portfolio inflows could relax financing constraints and thereby promote domestic innovation.

To address such concerns, we construct an instrument based on the idea of regional innovation spillover effects. Previous studies such as \cite{jaffe1993geographic} and \cite{audretsch1996r} highlight that innovation diffusion is region-specific due to geographical proximity and trade links. In addition, \cite{keller2002geographic}, \cite{moretti2004workers}, and \cite{bloom2013identifying} confirm that regional characteristics significantly influence technological spillovers. By contrast, domestic financial development and financial conditions are shaped by country-specific legal and institutional characteristics \citep{la1997legal,porta1998law,levine2023legal}.

\subsection{The Baseline Model}

Specifically, we utilize an interaction term as our regional push instrumental variable (IV), constructed as follows. Let $S_{r}$ denote the set of all countries in the region $r$. We separate the entire world into four regions: Asia, Europe, the Western Hemisphere, and Africa. We first identify the regional innovation leader $l_r$ in each region $r$ as the country $j$ with the highest absolute number of patent applications in a chosen base year $t_{1980}$, formally defined as
\begin{equation}\label{eq:leader_fpi}
l_{r} = \arg \max_{j \in S_{r}} \ \text{pat}_{j, t_{1980}}
\end{equation}
The regional leaders based on the equation above are Japan (Asia), Germany (Europe), the United States (Western Hemisphere), and South Africa (Africa). For each country $i$ in region $r$, we compute the average patent growth among all other countries in the same region except $i$, defined as
\begin{equation}
g_{r, t}^{-i} 
=
\frac{1}{|S_{r}| - 1}
\sum_{\substack{j \in S_{r}, j \neq i}}
\Delta\text{pat}_{j, t}.
\end{equation}
We then define $d_{i,l_r}$ as the geographical distance between the largest city of country $i$ and the regional leader $l_r$. These distances are calculated using the great circle formula, based on the latitudes and longitudes of the most important cities in each country \citep{head2002illusory}. Finally, the instrument for country $i$ in year $t$ is constructed as
\begin{equation}
z_{i, t} = \ln(d_{i,l_r}) \times g_{r, t}^{-i}.
\end{equation}
This instrument captures regional innovation spillover effects that influence domestic innovation activities while remaining plausibly exogenous to FPI. The validity stems from the fact that FPI typically does not exhibit regional characteristics, but is largely influenced by global push factors, market size, and informational advantage \citep{griffin2004daily,albuquerque2007international,andrade2010information,sarno2016drives}. In particular, while innovation can respond strongly to regional knowledge flows, equity and debt inflows in a country are strongly associated with other factors that are not directly influenced by the patenting activities of regional innovation leaders. Thus, the regional instrument affects FPI only through its impact on domestic innovation, which supports our instrument remains relevant for domestic innovation while satisfying the exclusion restriction for FPI. 

To construct a global push instrument, we take the ratio of the number of researchers in a country to the highest observed number of researchers in the dataset in any given year
\begin{equation}\label{eq:globiv}
    R_{i,t} = \frac{n_{i,t}}{\max_{j \in \mathcal{C}} n_{j,t}}
\end{equation}
where $\mathcal{C}$ denotes the set of all countries in the dataset.

To fully identify the causal effect of innovation on FPI, we further instrument the intensity of innovation with the capacity of a country's researchers scaled by the world's frontier each year. The exclusion restriction is credible after conditioning on financial development, trade openness, human capital, government size, inflation, real growth, banking crises, year, and regional fixed effects. Foreign investors respond primarily to verifiable, observable, and value-relevant signals of innovation rather than to inputs that are hard to interpret from abroad. An increase in domestic researcher headcount is a noisy signal for foreign investors, as it may reflect public hiring, reclassification, or projects with uncertain appropriability, and most importantly, more researchers do not necessarily mean a higher level of innovative output. Moreover, unlike patentable IP, it does not create protectable cash flow rights or reduce valuation uncertainty. In contrast, patent outputs are observable to global investors, certify novelty and forward-looking growth, and proxy for the likelihood that they will translate into scalable products and earnings. Hence, additional researchers should not directly attract equity inflow or debt inflow  unless they generate patentable output that is visible to, and credible for, international investors; local investors may react to headcount, but cross-border investors require patent-level evidence. This logic underpins our exclusion restriction: Frontier-scaled researcher capacity affects FPI only through its impact on patent production.

We employ a 2SLS estimator to estimate the causal impact of innovation on FPI. In the first stage, we regress current patent applications per million people on our constructed instruments, control variables, regional and time-fixed effects, excluding country-fixed effects since regional fixed effects are included in the first stage. The fitted values obtained represent the exogenous variation in innovation activity. In the second stage, we regress our measures of FPI on these fitted innovation values. This panel IV strategy is intended to isolate exogenous variation in innovation and support a causal interpretation of the innovation-FPI relationship. The weak-instrument diagnostics reported support instrument relevance and mitigate weak-identification concerns. Furthermore, the coefficient estimates on innovation remain robust and statistically significant across specifications. This consistency suggests that our strategy effectively captures the exogenous variation, thereby supporting the conclusions drawn from our causal inferences in the empirical analysis.

Our baseline second-stage specification is given by
\begin{equation}\label{eq:fpi_main}
	 \ln (y^s_{i,t}) = \beta_1 \ln (x_{i,t})+ \beta\mathbf{W}_{i,t} + \mu_t + \pi_r +\varepsilon_{i,t}
\end{equation}
where $y^s_{i,t}$ denotes equity inflow, debt inflow, the E/D  inflow ratio, or total FPI for country $i$ in year $t$, as defined in Table \ref{table:data_description}; $x_{i,t}$ is the number of patent applications per million people for country $i$ in year $t$; $\mathbf{W}_{i,t}$ denotes a vector of conditioning variables to control for other characteristics that may influence FPI including the years of schooling and returns to education, the banking crisis dummy, the measure of financial development, the general government consumption, the annual change of the consumer price index (CPI), the real GDP growth rate, and the sum of exports and imports of goods and services;
$\mu_t$ and $\pi_r$ are time-fixed effects and regional fixed effects; and $\varepsilon_{i,t}$ represents the regression residual, with robust standard errors \citep{white1980heteroskedasticity}. The first stage regression equation is given by 
\begin{equation}
    \ln (x_{i,t})=\phi_1z_{i,t} + \phi_2 R_{i,t} +\gamma\mathbf{W}_{i,t} + \theta_t +  \pi_r + \epsilon_{i,t}
\end{equation}
where $z_{i, t}$ serves as an instrument that proxies regional innovation spillovers that affect the country $i$ in year $t$; and \({R}_{i,t} \) represents a global push instrument based on (\ref{eq:globiv}).
Table \ref{table:base_fpi} reports the baseline IV estimates for the effect of innovation on FPI. 
Columns (1)–(2) present results for equity inflow with and without controls, and Columns (3)–(4) for debt inflow with and without controls.

The first-stage and weak-instrument diagnostics indicate strong instrument relevance. The regional push instruments enter negatively and significantly across all specifications because a greater distance to regional innovation leaders means lower exposure to their innovation capacity, whereas the global push instrument is positive and highly significant, indicating that both sources of exogenous variation load strongly on domestic patenting activity. Conventional first-stage F-statistics are very large, ranging from approximately 107 to 288, which is well above common relevance thresholds. Importantly, weak-instrument diagnostics are uniformly robust: the p-values of CLR, AR, and Wald tests are close to zero in every column, supporting inference that is robust to weak-identification concerns. Together, these results indicate that the instruments are highly relevant.

\begin{table}[ht]
        \centering
        \footnotesize
        \begin{threeparttable}
        \begin{tabular}{
            >{\raggedright\arraybackslash}p{2.5cm} 
            >{\centering\arraybackslash}m{2cm} 
            >{\centering\arraybackslash}m{2cm} 
            >{\centering\arraybackslash}m{2cm}  
            >{\centering\arraybackslash}m{2cm}  
        } 
             \toprule
             & \multicolumn{2}{c}{Equity inflow} & \multicolumn{2}{c}{Debt inflow} \\
             \cmidrule(lr){2-3} \cmidrule(lr){4-5}
             & (1) & (2) & (3) & (4) \\
             \midrule

            \underline{2nd Stage} &  &  &  &  \\

            Innovation & 1.156***  & 0.804***  & 0.717***  & 0.484***  \\
            &  (0.174) &  (0.082) &  (0.125) &  (0.067) \\
            
            \underline{1st Stage} &  &  &  &  \\

            Regional push & -0.153***  & -0.144*** & -0.170***  & -0.128***  \\
            & (0.044) &  (0.049) &  (0.037) &  (0.038) \\
            Global push & 3.287***  & 3.889*** & 3.462***  & 3.912***  \\
            &  (0.236) & (0.184) &  (0.215) &  (0.169) \\
           \underline{Weak IV Test} &  &  &  &  \\
            CLR  & 0.000 & 0.000 & 0.000 & 0.000 \\
            AR  & 0.000 & 0.000 & 0.000 & 0.000 \\
            Wald  & 0.000 & 0.000 & 0.000 & 0.000 \\ 
            1st F-statistic  & 106.779 & 243.076 & 142.190 & 288.285 \\
            \midrule
            Period & 1996-2021 & 1996-2021 & 1996-2021 & 1996-2021 \\
            Time FE & YES & YES & YES & YES \\
            Region FE & YES & YES & YES & YES \\
            Controls & YES & NO & YES & NO \\
            Obs. & 501 & 763 & 500 & 795 \\
            R-squared & 0.191 & 0.108 & 0.037 & 0.044 \\
           [1ex] 
           \bottomrule
        \end{tabular}
        \begin{tablenotes}
            \small
            \item The dependent variable is equity inflow or debt inflow. The endogenous variable is the number of patents per million people. Control variables are in Table \ref{table:data_description}. Note: *** \( p < 0.01 \), ** \( p < 0.05 \), * \( p < 0.1 \). Numbers in parentheses are robust standard errors. AR and Wald tests follow the procedures in \cite{olea2013robust}. Multiple IVs yield extra CLR statistics; see \cite{pflueger2015robust} for discussions of weak instrument tests in linear IV regressions and \cite{finlay2014weakiv10} for Stata implementations. $P$-values are reported for CLR, AR, and Wald tests.
        \end{tablenotes}
        \caption{Baseline Results}
        \label{table:base_fpi}
    \end{threeparttable}
    \end{table}

Innovation is positively associated with the scale of equity inflow. With controls, Column (1) shows a coefficient of 1.156, while without controls, Column (2) reports a coefficient of 0.804. Both estimates are statistically significant at the 1\% level. These coefficients imply that a 1\% increase in patents per million people is associated with approximately a 1.16\% increase in equity inflow when controls are included. Innovation is also positively associated with debt inflow. With controls, Column (3) reports a coefficient of 0.717, and without controls, Column (4) reports a coefficient of 0.484, again both significant at the 1\% level. Comparing across asset classes, the equity coefficients are larger than the corresponding debt coefficients in both specifications. This suggests a pronounced composition asymmetry: innovation attracts portfolio capital disproportionately more in the form of equity.
    
The pattern is consistent with the hypothesis that innovation, as a forward-looking signal of productivity and growth opportunities, attracts risk-bearing capital more strongly to equity than to debt. Debt inflow also responds positively, suggesting that innovative environments support broader external financing; however, equity inflow, with its higher liquidity and option-like payoff, appears more sensitive to shifts in expected returns associated with innovative activity. The similarity of signs and significance across specifications with and without controls, together with the uniformly strong identification diagnostics, indicates that these findings are not driven by specification choices.

In sum, the baseline evidence shows that (i) the instrument set passes strict relevance tests; (ii) innovation has a positive and statistically robust effect on both debt and equity inflows; and (iii) the magnitude is considerably larger for equity inflow, implying that innovation not only scales the overall foreign portfolio inflows but also shifts its composition toward equity. These results provide a clean, quantitative benchmark for the claim advanced in the paper: innovation helps explain why advanced, innovation-intensive economies are persistent net recipients of global portfolio capital, with the strongest channel operating through equity inflow.

\begin{table}[ht]
        \centering
        \footnotesize
        \begin{threeparttable}
        \begin{tabular}{
             >{\raggedright\arraybackslash}p{2.5cm} 
            >{\centering\arraybackslash}m{2cm} 
            >{\centering\arraybackslash}m{2cm} 
            >{\centering\arraybackslash}m{2cm}  
            >{\centering\arraybackslash}m{1.8cm}  
        } 
             \toprule
             & \multicolumn{2}{c}{E/D inflow ratio} & \multicolumn{2}{c}{Total FPI} \\
             \cmidrule(lr){2-3} \cmidrule(lr){4-5}
             & (1) & (2) & (3) & (4) \\
             \midrule

            \underline{2nd Stage} &  &  &  &  \\

            Innovation & 0.314** & 0.192** & 0.818*** & 0.577*** \\
            & (0.145) & (0.082) & (0.126) & (0.062) \\

            \underline{1st Stage} &  &  &  &  \\

            Regional push & -0.179*** & -0.174*** & -0.167*** & -0.131*** \\
            & (0.040) & (0.047) & (0.040) & (0.039) \\

            Global push & 3.448*** & 3.688*** & 3.398*** & 3.966*** \\
            & (0.223) & (0.178) & (0.216) & (0.169) \\

            \underline{Weak IV Test} &  &  &  &  \\
            CLR  & 0.031 & 0.018 & 0.000 & 0.000 \\
            AR  & 0.096 & 0.059 & 0.000 & 0.000 \\
            Wald  & 0.030 & 0.019 & 0.000 & 0.000 \\ 
            1st F-statistic  & 128.623 & 240.139 & 136.155 & 296.227 \\
            \midrule
            Period & 1996-2021 & 1996-2021 & 1996-2021 & 1996-2021 \\
            Time FE & YES & YES & YES & YES \\
            Region FE & YES & YES & YES & YES \\
            Controls & YES & NO & YES & NO \\
            Obs. & 439 & 668 & 529 & 843 \\
            R-squared & 0.231 & 0.133 & 0.066 & 0.093 \\
           [1ex] 
           \bottomrule
        \end{tabular}
        \begin{tablenotes}
            \small
            \item The dependent variable is the E/D inflow ratio or the total FPI. The endogenous variable is the number of patents per million people. Control variables are in Table \ref{table:data_description}. Note: *** \( p < 0.01 \), ** \( p < 0.05 \), * \( p < 0.1 \). Numbers in parentheses are robust standard errors. AR and Wald tests follow the procedures in \cite{olea2013robust}. Multiple IVs yield extra CLR statistics; see \cite{pflueger2015robust} for discussions of weak instrument tests in linear IV regressions and \cite{finlay2014weakiv10} for Stata implementations. $P$-values are reported for CLR, AR, and Wald tests.
        \end{tablenotes}
        \caption{Baseline Results: E/D Inflow Ratio and Total FPI}
        \label{table:based2}
    \end{threeparttable}
    \end{table}

Table \ref{table:based2} extends the baseline by replacing the dependent variable with (i) the E/D inflow ratio and (ii) total FPI, defined as equity inflow plus debt inflow. The first stage remains strong: the regional push is negative and significant, the global push is positive and highly significant, first-stage F-statistics are large, ranging from 128.623 to 296.227, and weak-IV tests generally reject at conventional levels. In the second stage, innovation confirms the shift in composition toward equity: the coefficient of the E/D inflow ratio with respect to patents per million people is 0.314 in Column (1), implying that a 1\% increase in patenting is associated with a 0.31\% higher E/D inflow ratio. Innovation also has a clear scaling effect on the portfolio inflows of countries: the coefficient of total FPI is 0.818 in Column (3), implying that a 1\% increase in patents per million people is associated with a 0.82\% rise in total FPI. This confirms that innovation both tilts foreign portfolio inflows toward equity and expands overall FPI, with equity-type capital responding more strongly to innovative opportunities.

\subsection{Proximity to Technological Frontier}

Innovation may not carry the same economic meaning across countries at very different stages of technological development. A patent in a country close to the world frontier may signal a technology with stronger commercial potential, greater relevance for global investors, and a clearer connection to future firm valuation. In contrast, a patent in a country far from the frontier may reflect technological catching-up or local adaptation, which may be important for domestic development but less immediately priced by international portfolio investors. Hence, the effect of innovation on FPI is expected to be stronger when the country is closer to the technological frontier.

Following \cite{acemoglu2006distance}, we construct an indicator, called proximity to technological frontier (PTF), to measure a country's productivity distance from the world technological frontier, thereby enabling us to identify whether the effect of innovation on FPI depends on the level of technological development.
To construct PTF, we take the ratio of a country's TFP to the highest observed TFP in the dataset in any given year
\begin{equation}
    \text{PTF}_{i,t} = \frac{TFP_{i,t}}{\max_{j \in \mathcal{C}} TFP_{j,t}}
\end{equation}
where  $\mathcal{C}$ denotes the set of all countries in the dataset. PTF  takes values in the range $(0,100]$ since it is multiplied by 100, where a value of 100 indicates that the country is on the world technological frontier, while lower values indicate that the country is lagging behind. This measure allows us to quantify how closely a country follows the most productive country over time.

We consider a regression with PTF interacting with the instrumented patent applications.  The second-stage equation is given by
\begin{equation}
\ln (y^s_{i,t}) = \omega_1 \ln (x_{i,t}) + \omega_2 \text{PTF}_{i,t} + \omega_3 \ln (x_{i,t}) \times \text{PTF}_{i,t} + \omega \mathbf{W}_{i,t} + \mu_t + \pi_r + \varepsilon_{i,t}.
\end{equation}

Table \ref{table:level_fpi} reports the estimates for $\omega_3$, the interaction effect of the instrumented patent applications with PTF. The interaction effect of innovation with PTF is strong for equity inflow. In Column (1), the coefficient of 1.722 is significant at the 1\% level. Thus, for a 1\% increase in patenting, a country with a one-unit increase in PTF experiences an additional 1.72\% increase in equity inflow relative to a country farther from the frontier. The weak-IV diagnostics are also reassuring for this column, with the CLR, AR, and Wald tests all rejecting at conventional levels, supporting the interpretation that technological proximity amplifies the response of equity inflow to innovation.

In contrast, the interaction for debt inflow in Column (2) is negative and insignificant. This suggests that the amplification effect associated with proximity to the frontier does not operate through debt inflow. Column (3) reports a similar compositional implication for the E/D inflow ratio: the coefficient is 1.572 and significant at the 5\% level, indicating that innovation tilts portfolio inflows toward equity as PTF improves. However, the weak-IV evidence for this column is more cautious: the AR test does not reject at conventional levels. For total FPI, defined as debt inflow plus equity inflow, the interaction in Column (4) is positive but statistically insignificant. Given the strong, precisely estimated effect on equity inflow and the positive but more cautiously supported effect on the E/D inflow ratio, this asymmetry indicates that as countries approach the frontier, innovation increasingly attracts risk-bearing equity inflow rather than debt inflow.
\begin{table}[ht]
    \centering
    \footnotesize
    \begin{threeparttable}
    \begin{tabular}{
        >{\raggedright\arraybackslash}p{2.5cm} 
        >{\centering\arraybackslash}m{2.1cm} 
        >{\centering\arraybackslash}m{2.1cm}
        >{\centering\arraybackslash}m{2.6cm}
        >{\centering\arraybackslash}m{2.1cm}
    } 
         \toprule
         & Equity inflow & Debt inflow & E/D inflow ratio & Total FPI \\
         \cmidrule(lr){2-2} \cmidrule(lr){3-3} \cmidrule(lr){4-4} \cmidrule(lr){5-5}
         & (1) & (2) & (3) & (4) \\
         \midrule

        \underline{2nd Stage} &  &  &  &  \\

        Patents*PTF & 
        1.722***  & -0.188 & 1.572** & 0.678 \\
         & 
        (0.487)  & (0.610)  & (0.660) & (0.463) \\

        \underline{Weak IV Test} &  &  &  &  \\
        CLR  & 0.000 & 0.006 & 0.045 & 0.002 \\
        AR   & 0.000 & 0.031 & 0.149 & 0.009 \\
        Wald & 0.000 & 0.028 & 0.055 & 0.009 \\

        \midrule
        Period   & 1996-2021 & 1996-2021 & 1996-2021 & 1996-2021 \\
        Time FE  & YES & YES & YES & YES \\
        Region FE  & YES & YES & YES & YES \\
        Controls & YES & YES & YES & YES \\
        Obs. & 496 & 497 & 436 & 525 \\
        R-squared& 0.341 & 0.139 & 0.235 & 0.282 \\
       [1ex] 
       \bottomrule
    \end{tabular}
    \begin{tablenotes}
        \footnotesize
        \item The dependent variable is either equity inflow, debt inflow, E/D inflow ratio, or total FPI. The endogenous variable is the number of patent applications instrumented with the regional innovation spillover instrument and regional dummies. Control variables are in Table \ref{table:data_description}. Note: *** \( p < 0.01 \), ** \( p < 0.05 \), * \( p < 0.1 \). Numbers in parentheses are robust standard errors. AR and Wald tests follow \cite{olea2013robust}. With multiple IVs, we report CLR statistics; see \cite{pflueger2015robust} for weak-IV tests in linear IV regressions and \cite{finlay2014weakiv10} for Stata implementations. $P$-values are reported for CLR, AR, and Wald tests. Because F-statistics may be unreliable with more than one endogenous regressor, we rely on weak-IV tests for interaction regressions.
    \end{tablenotes}
    \caption{Interaction of Innovation with Technological Development}
    \label{table:level_fpi}
\end{threeparttable}
\end{table}

In general, across specifications with regional and year-fixed effects and controls, proximity to the frontier amplifies the impact of innovation on foreign portfolio flows, most clearly on  equity inflow and on the E/D inflow ratio, while the effect on debt inflow alone is weak. Thus, advancing toward the technological frontier mainly strengthens the equity response and provides suggestive evidence of a tilt toward equity, consistent with foreign investors responding more to innovative, growth-oriented opportunities in frontier economies.

\subsection{Institutional Quality}\label{fpi:inst}

The preceding analysis demonstrates that the impact of innovation on FPI intensifies as countries approach the world technological frontier, reflecting variations in technological capacity and absorptive capacity across economies. However, PTF alone may not be sufficient to fully understand how innovation attracts FPI, especially how this relationship is heterogeneous depending on asset classes.  Even countries at similar levels of innovation capability can exhibit different results in attracting FPI if institutional quality differs. Institutions influence the effectiveness and confidence with which innovation is translated into economic activity and whether they can adapt to support foreign financing. Strong institutions promote legal certainty, regulatory quality, and transparency, all of which are crucial for fostering international investor confidence and enabling financial markets to respond to innovative activities.

Therefore, this section extends our empirical framework to examine whether institutional quality moderates the impact of innovation on FPI. We analyze six key institutional dimensions following the Worldwide Governance Indicators (WGIs): regulatory quality, voice and accountability, corruption control, rule of law, political stability, and government effectiveness. Our second-stage empirical specification remains consistent with the approach used in the previous section
\begin{equation}
\ln (y^s_{i,t}) = \sigma_1 \ln (x_{i,t}) + \sigma_2 q_{i,t} + \sigma_3 \ln (x_{i,t}) \times q_{i,t} + \sigma \mathbf{W}_{i,t} + \mu_t + \pi_r + \varepsilon_{i,t}
\end{equation}
where $q_{i,t}$ represents one of six institutional quality measures for country $i$ in year $t$. 

\begin{table}[ht]   
    \centering
    \footnotesize
    \begin{threeparttable}
    \begin{tabular}{
        >{\raggedright\arraybackslash}p{3cm} 
        >{\centering\arraybackslash}m{1.2cm} 
        >{\centering\arraybackslash}m{1.2cm} 
        >{\centering\arraybackslash}m{1.2cm} 
        >{\centering\arraybackslash}m{1.2cm} 
        >{\centering\arraybackslash}m{1.2cm} 
        >{\centering\arraybackslash}m{1.2cm} 
    } 
         \toprule
         & \multicolumn{2}{c}{Regulatory quality} 
         & \multicolumn{2}{c}{Political stability}
         & \multicolumn{2}{c}{Voice accountability} \\
         \cmidrule(lr){2-3} 
         \cmidrule(lr){4-5} 
         \cmidrule(lr){6-7}
         & Equity inflow & Debt inflow & Equity inflow & Debt inflow & Equity inflow & Debt inflow \\
         \midrule

        \underline{2nd Stage} &  &  &  &  &  &  \\

        Patents*Institutions 
        & 0.321** & 0.085 
        & 0.360*** & 0.167* 
        & 0.541*** & 0.351** \\
        & (0.139) & (0.125)
        & (0.104) & (0.096)
        & (0.166) & (0.145) \\

        \underline{Weak IV Test} &  &  &  &  &  &  \\
        CLR  
        & 0.001 & 0.823 & 0.000 & 0.000 & 0.000 & 0.000 \\
        AR  
        & 0.000 & 0.106 & 0.000 & 0.000 & 0.000 & 0.002 \\
        Wald  
        & 0.001 & 0.539 & 0.000 & 0.000 & 0.000 & 0.001 \\

        \midrule
        Period   & 1996-2021 & 1996-2021 & 1996-2021 & 1996-2021 & 1996-2021 & 1996-2021 \\
        Time FE  & YES & YES & YES & YES & YES & YES \\
        Region FE & YES & YES & YES & YES & YES & YES \\
        Controls & YES & YES & YES & YES & YES & YES \\
        Obs.     & 443 & 449 & 451 & 452 & 451 & 452 \\
        R-squared & 0.381 & 0.253 & 0.313 & 0.119 & 0.370 & 0.172 \\
       [1ex] 
       \bottomrule
    \end{tabular}
    \begin{tablenotes}
        \small
        \item The dependent variable is equity inflow or debt inflow. The endogenous variable is the number of patents per million people. Control variables are in Table \ref{table:data_description}. Note: *** \( p < 0.01 \), ** \( p < 0.05 \), * \( p < 0.1 \). Numbers in parentheses are robust standard errors. AR and Wald tests follow the procedures in \cite{olea2013robust}. Multiple IVs yield extra CLR statistics; see \cite{pflueger2015robust} for discussions of weak instrument tests in linear IV regressions and \cite{finlay2014weakiv10} for Stata implementations. $P$-values are reported for CLR, AR, and Wald tests. F-statistics may have questionable accuracy in regressions with more than one endogenous regressor; thus, we rely on weak IV test results for our interaction regressions.
    \end{tablenotes}
    \caption{Regulation, Politics, and Voice Accountability Channels}
    \label{table:inst}
\end{threeparttable}
\end{table}

Table \ref{table:inst} shows that across three institutional quality dimensions, the interaction effect is large and positive for equity inflow. The interaction effect with regulatory quality is 0.321 for equity inflow, while the debt inflow counterpart is 0.085 and statistically insignificant. The interaction effect with political stability is 0.360 for equity inflow, compared to 0.167 for debt inflow. The interaction effect with voice and accountability is 0.541 for equity inflow, compared with 0.351 for debt inflow. These results indicate that a one-unit improvement in the institutional index for these three indicators raises the innovation coefficient of equity inflow by about 0.32 to 0.54. The corresponding change for debt inflow is generally smaller, although it becomes statistically significant for political stability and voice and accountability. In other words, as institutions strengthen, an increase in patenting translates into a larger increase in equity inflow, while the amplification for debt inflow is weaker and less uniform across the above three institutional dimensions.

\begin{table}[ht]   
   \centering
    \footnotesize
    \begin{threeparttable}
    \begin{tabular}{
        >{\raggedright\arraybackslash}p{3cm} 
        >{\centering\arraybackslash}m{1.2cm} 
        >{\centering\arraybackslash}m{1.2cm} 
        >{\centering\arraybackslash}m{1.2cm} 
        >{\centering\arraybackslash}m{1.2cm} 
        >{\centering\arraybackslash}m{1.2cm} 
        >{\centering\arraybackslash}m{1.2cm} 
    } 
         \toprule
         & \multicolumn{2}{c}{Rule of law} 
         & \multicolumn{2}{c}{Corruption controls} 
         & \multicolumn{2}{c}{Gov. effectiveness} \\
         \cmidrule(lr){2-3} 
         \cmidrule(lr){4-5} 
         \cmidrule(lr){6-7}
         & Equity inflow & Debt inflow & Equity inflow & Debt inflow & Equity inflow & Debt inflow\\
         \midrule

        \underline{2nd Stage} &  &  &  &  &  &  \\

        Patents*Institutions  
        & 0.203* & -0.014
        & 0.107 & -0.009
        & 0.025 & -0.183 \\
        & (0.121) & (0.116)
        & (0.108) & (0.096)
        & (0.135) & (0.150) \\

        \underline{Weak IV Test} &  &  &  &  &  & \\
        CLR  
        & 0.115 & 0.490 & 0.031 & 0.095 & 0.007 & 0.009 \\
        AR  
        & 0.010 & 0.440 & 0.006 & 0.190 & 0.005 & 0.028 \\
        Wald  
        & 0.041 & 0.518 & 0.027 & 0.140 & 0.003 & 0.023 \\

        \midrule
        Period & 1996-2021 & 1996-2021 & 1996-2021 & 1996-2021 & 1996-2021 & 1996-2021 \\
        Time FE & YES & YES & YES & YES & YES & YES \\
        Region FE & YES & YES & YES & YES & YES & YES \\
        Controls & YES & YES & YES & YES & YES & YES \\
        Obs. & 451 & 452 & 451 & 452 & 451 & 452 \\
        R-squared & 0.408 & 0.200 & 0.401 & 0.163 & 0.385 & 0.149 \\
       [1ex] 
       \bottomrule
    \end{tabular}
    \begin{tablenotes}
        \small
        \item The dependent variable is equity inflow or debt inflow. The endogenous variable is the number of patents per million people. Control variables are in Table \ref{table:data_description}.  Note: *** \( p < 0.01 \), ** \( p < 0.05 \), * \( p < 0.1 \). Numbers in parentheses are robust standard errors. AR and Wald tests follow the procedures in \cite{olea2013robust}. Multiple IVs yield extra CLR statistics; see \cite{pflueger2015robust} for discussions of weak instrument tests in linear IV regressions and \cite{finlay2014weakiv10} for Stata implementations. $P$-values are reported for CLR, AR, and Wald tests. F-statistics may have questionable accuracy in regressions with more than one endogenous regressor; thus, we rely on weak IV test results for our interaction regressions.
    \end{tablenotes}
    \caption{Law, Corruption, and Government Channels}
     \label{table:insthree}
\end{threeparttable}

\end{table}

Identification diagnostics generally support this pattern, but with an important qualification. The equity inflow columns pass the weak-IV checks, with CLR, AR, and Wald tests rejecting at conventional levels across the three institutional dimensions. For debt inflow, the diagnostics are also strong for political stability and voice and accountability, but the regulatory quality specification shows no support. Overall, the evidence suggests a composition effect: higher institutional quality enhances the ability of innovation to attract risk-bearing equity inflow more strongly than debt inflow. This pattern aligns with the idea that, in stronger institutional environments, innovation credibly signals high, but uncertain, growth opportunities that international investors are more willing to finance through equity rather than fixed-income claims.

\section{Development-Ladder Process and Threshold Effect}\label{sec:hete}

\begin{figure}[ht!]
\begin{center}
	\includegraphics[width=1.0\textwidth]{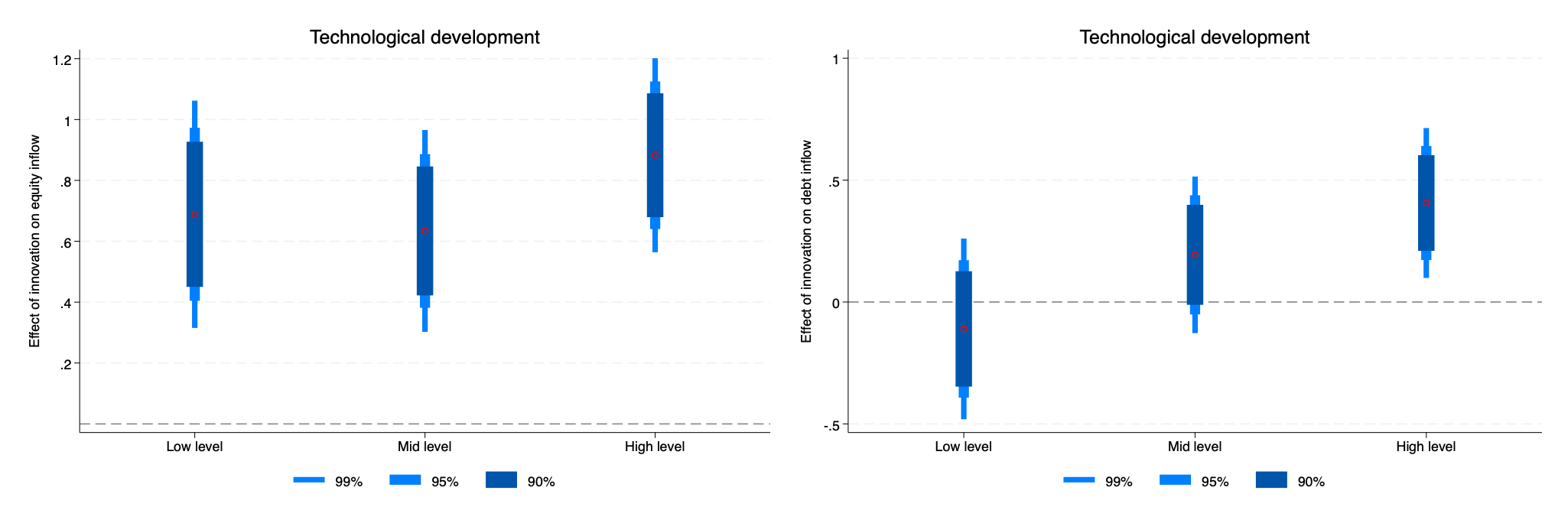}
\caption{Interaction of Innovation with Three PTF Groups}
\label{figure:nonptf}
\end{center}
\end{figure}

In this section, we test whether the effect of innovation on FPI is nonlinear, as the linear mean effect masks economically plausible heterogeneity across institutional environments. The average coefficient of the interaction term with innovation is often small or insignificant, as shown in Table \ref{table:insthree}, which could be due to offsetting effects: the effect of innovation on FPI may be heterogeneous so that it becomes only significant when institutional quality exceeds a certain threshold. Therefore, we estimate a grouped interaction model that allows the coefficients of innovation to vary with terciles of institutional quality.  In particular, we estimate 
\begin{equation}
\begin{split}
    \ln (y^s_{i,t}) = \beta\,\ln (x_{i,t})
  + \sum_{k=1}^{3}\theta_k\,\big(q_{i,t}\,\mathbb{I}(q_{i,t}\in G_k)\big) + \sum_{k=1}^{3}\lambda_k\,\big(\ln (x_{i,t})\, q_{i,t}\,\mathbb{I}(q_{i,t}\in G_k)\big)
  \\+ \gamma \mathbf W_{i,t} + \delta_t +\pi_r + \varepsilon_{i,t}. 
\end{split}
\end{equation}
Figure \ref{figure:nonptf} examines heterogeneity by splitting PTF values into tertiles and re-estimating the innovation coefficient of equity inflow (left panel) and debt inflow (right panel). The estimates are obtained from the same IV specification used in the baseline, with innovation instrumented, region and year effects included, and controls included. The dashed horizontal line represents zero; the bars indicate 99, 95, and 90 percent confidence intervals.

For debt inflow, the magnitude and significance of the effect of innovation depends on the level of PTF. In the low-PTF tertile, the point estimate is negative, and the confidence intervals cluster around zero, indicating that additional patenting does not translate into higher debt inflow and may even be associated with reduced debt inflow. Moving to the middle tertile, the estimate turns positive and insignificant, and in the top tertile, the effect is positive and significantly larger. Thus, innovation begins to attract debt only once countries reach the high technology development group.

The equity panel shows a stronger effect and development-ladder process. Innovation is associated with positive equity inflow even in the lowest tertile, of comparable magnitude in the middle tertile, and the effect rises substantially in the highest tertile, with narrow bands that exclude zero. Thus, the tertile results confirm the composition asymmetry documented in the tables in Section \ref{fpi:inst}: innovation consistently draws in equity inflow, and as countries approach the frontier, the responsiveness of portfolio inflow increases. However, foreign debt investors only view innovation as a credible signal of stable growth opportunities in the highest-PTF environment.

\begin{figure}[ht!]
\begin{center}
	\includegraphics[width=1.0\textwidth]{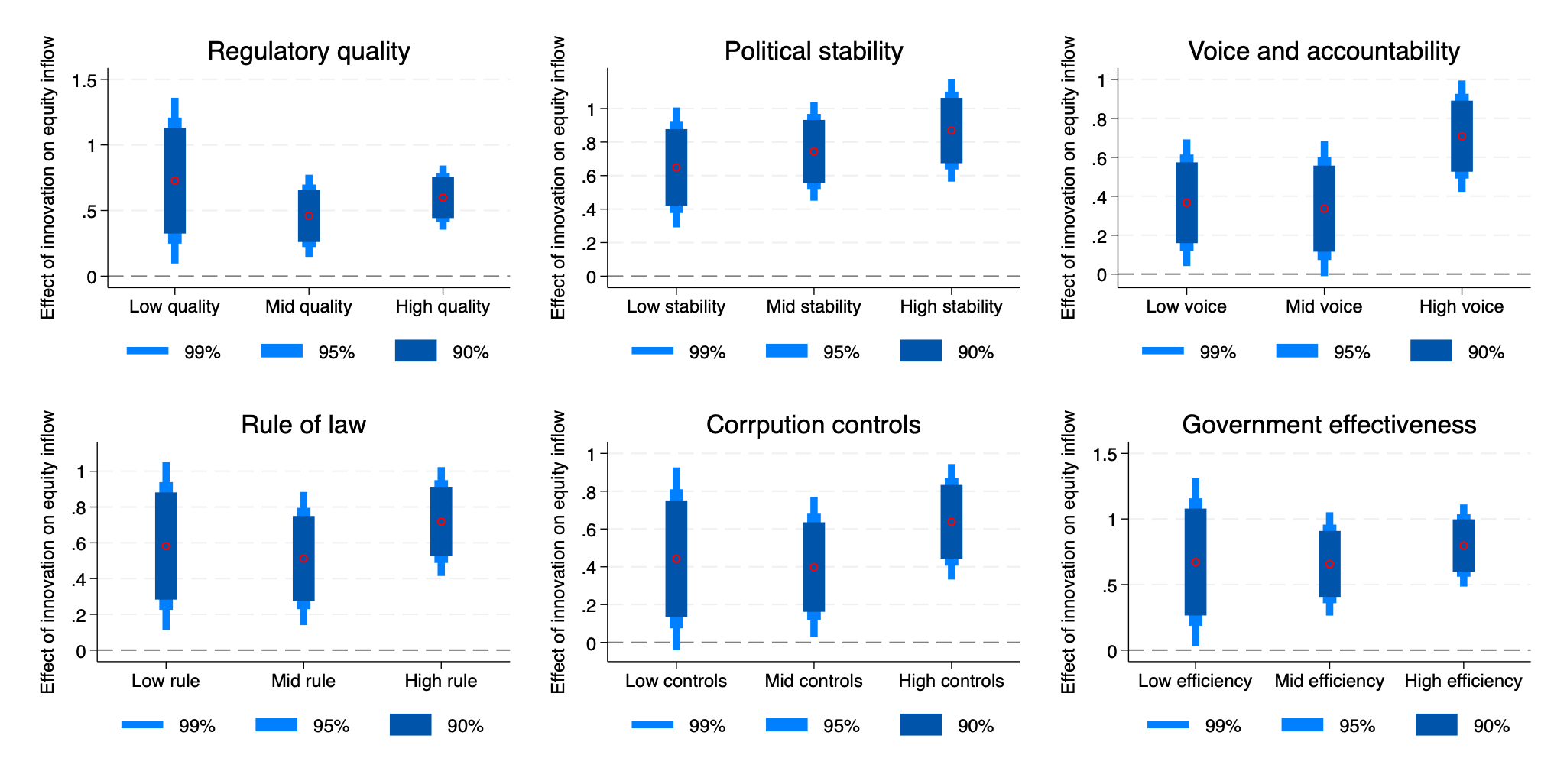}
\caption{Interaction of Innovation with Three Institutional Quality Groups for Equity Inflow}
\label{figure:non_inst_fe}
\end{center}
\end{figure}

\begin{figure}[ht!]
\begin{center}
	\includegraphics[width=1.0\textwidth]{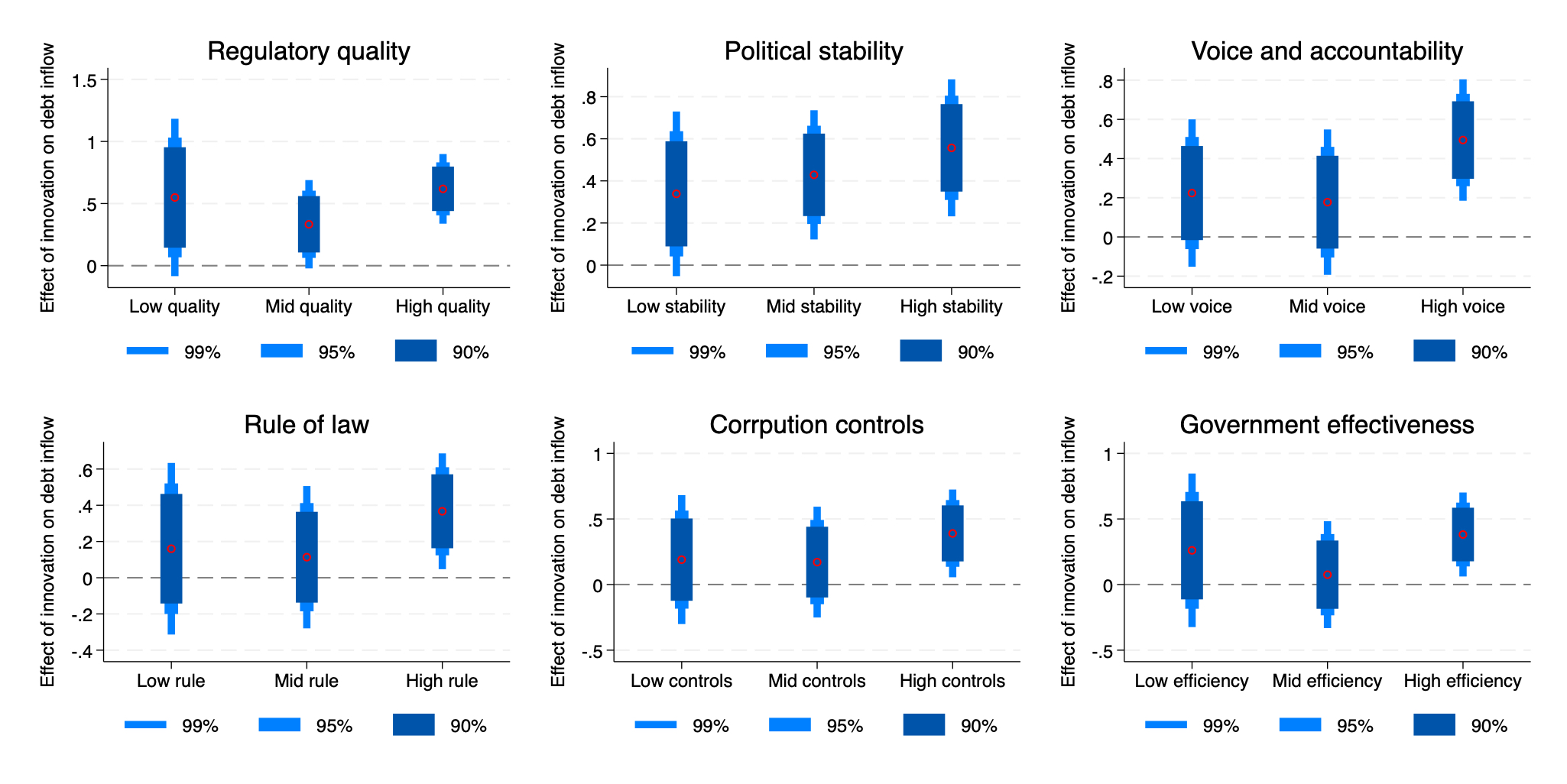}
\caption{Interaction of Innovation with Three Institutional Quality Groups for Debt Inflow}
\label{figure:non_inst_fd}
\end{center}
\end{figure}

Figures \ref{figure:non_inst_fe} and \ref{figure:non_inst_fd} report grouped IV estimates that split each of the six WGIs: regulatory quality, rule of law, political stability, voice and accountability, corruption control, and government effectiveness, into terciles. Within each tercile we estimate the effect of innovation on FPI. Together, the panels reveal a clear institutional gradient and a pronounced asymmetry between equity  inflow and debt inflow.

The equity inflow panels display a nearly monotonic trend. Even in the low-quality tercile, innovation is associated with positive equity inflow in all indicators, and the magnitude rises systematically as institutional quality improves. In the top tercile, the effects are large and more precisely estimated for all six indicators. The gradient is particularly steep for political stability, voice and accountability, rule of law, control of corruption, and government effectiveness, consistent with the view that equity investors respond to credible policy frameworks and market-friendly environments that enhance the payoffs to innovation and facilitate entry, listing, and trading.

The debt inflow panels show that innovation translates into higher external borrowing only at the very top of the institutional distribution for most of the indicators. In the low and mid-quality terciles, the point estimates are small, and the confidence bands overlap with zero. By contrast, in the high-quality tercile, the effect becomes positive and statistically distinct from zero across voice and accountability, rule of law, corruption control, and government effectiveness. These patterns suggest that the debt inflow channel is activated mainly when institutional quality reaches a high threshold, when the quality of democracy, contract enforcement, anti-corruption, and bureaucratic capacity collectively reduce default and rollover risk to levels compatible with sustained foreign lending.

This divergence between panels points to a composition asymmetry: innovation attracts equity inflow in relatively strong institutional environments, while the debt inflow response requires institutions that are exceptionally strong. A plausible explanation is that equity finance values growth options and can be redeployed or exited in the face of shocks, so moderate institutional strength suffices for innovation signals to draw equity inflow. Debt inflow, on the contrary, depends on enforceability and sovereign capacity to commit, because international debt investors seek stable returns and require enforceable repayment claims. Thus, innovation increases borrowing only once very high thresholds are met.

In summary, as institutions strengthen from low to middle levels, innovation mainly attracts equity inflow. Only after countries approach the top tercile of institutional quality does innovation bring in debt inflow. Reforms that raise the credibility and effectiveness of the state, particularly improvements in the quality of democracy, law enforcement, corruption control, and administrative capacity, therefore not only amplify the overall effect of innovation on FPI, but also determine whether the capital flow is in the form of equity early on or in the form of both equity and debt at later stages.

\section{Risk-Taking Environment}\label{sec:further_fpi}

Innovation creates investment opportunities with high expected returns, but these opportunities are also associated with greater uncertainty. Whether foreign investors respond to domestic innovation therefore depends not only on the existence of innovative activity, but also on the extent to which the host country allows international investors to bear, price, and manage risk. A more open capital account lowers barriers to cross-border portfolio adjustment and makes it easier for foreign investors to enter and exit domestic securities markets. Macroprudential policy shapes the risk-taking environment of the domestic financial system by affecting leverage, credit expansion, and the transmission of financial shocks. Legal origin, in turn, captures deeper institutional differences in investor protection, disclosure rules, and the enforceability of financial claims. These factors jointly determine whether innovation can be transformed into investable risk-bearing opportunities for foreign portfolio investors. We therefore examine financial openness, macroprudential policy, and legal origin as complementary dimensions of the risk-taking environment.

\subsection{Financial Openness}

Financial openness moderates the impact of domestic innovation on FPI, with equity inflow responding more strongly in more open economies.\footnote{The data on financial openness is obtained from \cite{chinn2008new}.} The empirical results obtained in Table \ref{table:fpi_finop} are consistent with evidence that liberalized capital accounts and reduced capital controls lower transaction costs and informational frictions, facilitating faster and larger equity inflow \citep{henry2000stock,lane2007external,forbes2015capital}. In contrast, debt inflow is comparatively less sensitive to openness, likely because foreign creditors require enforceable contracts and stronger institutional backups to manage default risk.

\begin{table}[ht]
    \centering
    \footnotesize
    \begin{threeparttable}
    \begin{tabular}{
        >{\raggedright\arraybackslash}p{3cm} 
        >{\centering\arraybackslash}m{2.2cm} 
        >{\centering\arraybackslash}m{2.2cm}
        >{\centering\arraybackslash}m{3.2cm}
        >{\centering\arraybackslash}m{2.2cm}
    } 
         \toprule
         & Equity inflow & Debt inflow & E/D inflow ratio & Total FPI \\
         \cmidrule(lr){2-2} \cmidrule(lr){3-3} \cmidrule(lr){4-4} \cmidrule(lr){5-5}
         & (1) & (2) & (3) & (4) \\
         \midrule

        \underline{2nd Stage} &  &  &  &  \\

        Patents*Financial Openness & 
        0.134***  & -0.067  & 0.194*** & -0.042 \\
         & 
        (0.039)  & (0.046)  & (0.059) & (0.041) \\

        \underline{Weak IV Test} &  &  &  &  \\
        CLR  & 0.000 & 0.005 & 0.002 & 0.001 \\
        AR   & 0.000 & 0.004 & 0.001 & 0.001 \\
        Wald & 0.000 & 0.003 & 0.004 & 0.000 \\

        \midrule
        Period   & 1996--2021 & 1996--2021 & 1996--2021 & 1996--2021 \\
        Time FE  & YES & YES & YES & YES \\
        Region FE  & YES & YES & YES & YES \\
        Controls & YES & YES & YES & YES \\
        Obs. & 644 & 651 & 547 & 695 \\
        R-squared& 0.324 & 0.097 & 0.143 & 0.247 \\
       [1ex] 
       \bottomrule
    \end{tabular}
    \begin{tablenotes}
        \footnotesize
        \item The dependent variable is either equity inflow, debt inflow, E/D inflow ratio, or total FPI. The endogenous variables are innovation and its interaction with financial openness, instrumented with regional innovation spillovers and their interactions with financial openness. Control variables are listed in Table \ref{table:data_description}. Note: *** \( p < 0.01 \), ** \( p < 0.05 \), * \( p < 0.1 \). Numbers in parentheses are robust standard errors. AR and Wald tests follow \cite{olea2013robust}. With multiple endogenous regressors, we report CLR statistics; see \cite{pflueger2015robust} for weak-IV tests in linear IV regressions and \cite{finlay2014weakiv10} for Stata implementations. $P$-values are reported for CLR, AR, and Wald tests. Because first-stage F-statistics may be unreliable with more than one endogenous regressor, we rely on weak-IV tests for interaction regressions.
    \end{tablenotes}
    \caption{Interaction of Innovation with Financial Openness}
    \label{table:fpi_finop}
\end{threeparttable}
\end{table}

Financial openness reduces barriers to cross-border investment, allowing international investors to capture returns from domestic innovation opportunities more efficiently. For equity inflow, which is liquid and relatively easier to trade, reduced regulatory constraints and capital controls enhance responsiveness. Debt inflow, which is subject to legal enforceability and creditor rights, remains constrained by institutional and legal limitations even in open economies. These findings complement previous studies emphasizing that capital account liberalization boosts total inflows \citep{bekaert2011financial}, but extend the literature by demonstrating that innovation is a critical driver of equity inflow under liberalization.

\subsection{Macroprudential Policy}

The effects of innovation on FPI are also sensitive to macroprudential regulation, proxied here by loan-to-value (LTV) ratios.\footnote{The data is collected from the Integrated Macroprudential Policy (iMaPP) Database.} In Table \ref{table:fpi_prud}, since higher LTV indicates weaker macroprudential constraints, the positive coefficients suggest that weaker regulation amplifies the impact of innovation on both equity inflow and debt inflow, with the stronger and more statistically significant effect for equity inflow. Weaker macroprudential measures increase the leverage and capacity of domestic financial systems, creating opportunities for foreign investors to engage with domestic markets. International investors, benefiting from liquidity and flexibility, can more readily exploit innovation signals when domestic regulation allows higher leverage and greater risk-taking. Debt inflow, however, requires formal creditor protections and institutional safeguards, so its responsiveness is comparatively limited under weaker regulation. This is consistent with literature emphasizing that macroprudential policies can moderate cross-border capital flows \citep{bruno2015capital,bergant2024dampening}.

\begin{table}[ht]
    \centering
    \footnotesize
    \begin{threeparttable}
    \begin{tabular}{
        >{\raggedright\arraybackslash}p{2.5cm} 
        >{\centering\arraybackslash}m{2.1cm} 
        >{\centering\arraybackslash}m{2cm}
        >{\centering\arraybackslash}m{3cm}
        >{\centering\arraybackslash}m{2cm}
    } 
         \toprule
         & Equity inflow & Debt inflow & E/D inflow ratio & Total FPI \\
         \cmidrule(lr){2-2} \cmidrule(lr){3-3} \cmidrule(lr){4-4} \cmidrule(lr){5-5}
         & (1) & (2) & (3) & (4) \\
         \midrule

        \underline{2nd Stage} &  &  &  &  \\

        Patents*LTV & 
        0.013**  & 0.010*  & 0.004 & 0.017*** \\
         & 
        (0.006)  & (0.006)  & (0.006) & (0.005) \\

        \underline{Weak IV Test} &  &  &  &  \\
        CLR  & 0.000 & 0.000 & 0.464 & 0.000 \\
        AR   & 0.000 & 0.000 & 0.651 & 0.000 \\
        Wald & 0.000 & 0.000 & 0.288 & 0.000 \\

        \midrule
        Period   & 1996--2021 & 1996--2021 & 1996--2021 & 1996--2021 \\
        Time FE  & YES & YES & YES & YES \\
        Region FE  & YES & YES & YES & YES \\
        Controls & YES & YES & YES & YES \\
        Obs.     & 574 & 607 & 509 & 632 \\ 
        R-squared& 0.079 & 0.688 & 0.208 & 0.669 \\  
       [1ex] 
       \bottomrule
    \end{tabular}
    \begin{tablenotes}
        \footnotesize
        \item The dependent variable is either equity inflow, debt inflow, E/D inflow ratio, or total FPI. The key endogenous interaction term is innovation interacted with average LTV, instrumented with regional innovation spillovers and their interactions with LTV. Control variables are listed in Table \ref{table:data_description}. Note: *** \( p < 0.01 \), ** \( p < 0.05 \), * \( p < 0.1 \). Numbers in parentheses are robust standard errors. AR and Wald tests follow \cite{olea2013robust}. With multiple endogenous regressors, we report CLR statistics; see \cite{pflueger2015robust} for weak-IV tests in linear IV regressions and \cite{finlay2014weakiv10} for Stata implementations. $P$-values are reported for CLR, AR, and Wald tests.
    \end{tablenotes}
    \caption{Interaction of Innovation with Macroprudential Policy}
    \label{table:fpi_prud}
\end{threeparttable}
\end{table}

Overall, our findings suggest that countries with higher LTV ratios, implying looser lending restrictions, experience stronger FPI responses to domestic innovation, particularly in equity markets. This has important policy implications: macroprudential regulations can shape the composition of foreign capital, and weaker constraints can amplify the equity inflow response to innovation while leaving debt inflow less affected.

\subsection{Legal Origins}\label{legal}

The results in Table \ref{table:fpi_legal} indicate substantial heterogeneity in innovation‐driven FPI across legal systems. Interaction coefficients are particularly strong in the UK, French, and Nordic legal origins for both equity inflow and debt inflow, whereas they are insignificant in German and Socialist legal origins for debt inflow and insignificant in Socialist legal origins for equity inflow. This pattern is consistent with the legal origins literature, which emphasizes that legal systems shape investor protection, disclosure requirements, and the enforceability of financial claims \citep{la1997legal,porta1998law,djankov2008law}. Legal systems with stronger investor protection provide clearer shareholder and creditor rights and rely heavily on private enforcement, which may make foreign investors more responsive to forward-looking information such as innovation activity.

The strong effect in French legal-origin countries should be interpreted cautiously. It may reflect the interaction between centralized regulation, disclosure requirements, and the way innovation signals are incorporated into financial markets. Although French civil law is often characterized by heavier state intervention and weaker private enforcement relative to common law, it features highly centralized regulatory oversight, extensive disclosure requirements, and strong administrative capacity, all of which can transform innovation outcomes into more credible and interpretable signals for foreign investors. Many French-origin economies—such as France, Belgium, Spain, and several Latin American countries—operate under regulatory regimes where innovation is closely monitored, certified, or supported by the state. 

\begin{table}[ht]
    \centering
    \footnotesize
    \makebox[\textwidth][c]{%
    \begin{threeparttable}
    \begin{tabular}{
        >{\raggedright\arraybackslash}p{2.3cm}
        >{\centering\arraybackslash}m{1.cm}
        >{\centering\arraybackslash}m{1.cm}
        >{\centering\arraybackslash}m{1.cm}
        >{\centering\arraybackslash}m{1.cm}
        >{\centering\arraybackslash}m{1.cm}
        >{\centering\arraybackslash}m{1.cm}
        >{\centering\arraybackslash}m{1.cm}
        >{\centering\arraybackslash}m{1.cm}
        >{\centering\arraybackslash}m{1.cm}
        >{\centering\arraybackslash}m{1.cm}
    }
         \toprule
         & \multicolumn{5}{c}{Equity inflow} & \multicolumn{5}{c}{Debt inflow} \\
         \cmidrule(lr){2-6} \cmidrule(lr){7-11}
         & UK & French & Social & German & Nordic & UK & French & Social & German & Nordic \\
         & (1) & (2) & (3) & (4) & (5) & (6) & (7) & (8) & (9) & (10) \\
         \midrule

        2nd Stage &  &  &  &  &  &  &  &  &  &  \\

        Interaction
        & 0.563***
        & 0.858***
        & 0.236
        & 0.364***
        & 0.520***
        & 0.437**
        & 0.568***
        & 0.236
        & 0.204
        & 0.442*** \\
        &
        (0.186)
        & (0.179)
        & (0.200)
        & (0.134)
        & (0.186)
        & (0.171)
        & (0.142)
        & (0.189)
        & (0.134)
        & (0.161) \\

        Weak IV Test &  &  &  &  &  &  &  &  &  &  \\

        CLR
        & 0.000 & 0.000 & 0.000 & 0.000 & 0.000
        & 0.000 & 0.000 & 0.000 & 0.000 & 0.000 \\

        AR
        & 0.000 & 0.000 & 0.000 & 0.000 & 0.000
        & 0.000 & 0.000 & 0.000 & 0.000 & 0.000 \\

        Wald
        & 0.000 & 0.000 & 0.000 & 0.000 & 0.000
        & 0.000 & 0.000 & 0.000 & 0.000 & 0.000 \\

        1st F-statistic
        & 63.21 & 63.21 & 63.21 & 63.21 & 63.21
        & 57.11 & 57.11 & 57.11 & 57.11 & 57.11 \\
        \midrule

        Period
        & 1996-2021 & 1996-2021 & 1996-2021 & 1996-2021 & 1996-2021
        & 1996-2021 & 1996-2021 & 1996-2021 & 1996-2021 & 1996-2021 \\

        Region FE
        & YES & YES & YES & YES & YES
        & YES & YES & YES & YES & YES \\

        Time FE
        & YES & YES & YES & YES & YES
        & YES & YES & YES & YES & YES \\

        Controls
        & YES & YES & YES & YES & YES
        & YES & YES & YES & YES & YES \\

        Obs.
        & 481 & 481 & 481 & 481 & 481
        & 483 & 483 & 483 & 483 & 483 \\

        R-squared
        & 0.441 & 0.441 & 0.441 & 0.441 & 0.441
        & 0.167 & 0.167 & 0.167 & 0.167 & 0.167 \\
       \bottomrule
    \end{tabular}
    \begin{tablenotes}
        \small
        \item The dependent variable is either equity inflow or  debt inflow as a percentage of GDP. The endogenous variables are legal-origin-specific innovation measures. Note: *** \( p < 0.01 \), ** \( p < 0.05 \), * \( p < 0.1 \). Numbers in parentheses are robust standard errors. Weak IV tests are the CLR, AR, and Wald tests following \cite{olea2013robust} and \cite{pflueger2015robust}.
    \end{tablenotes}
    \caption{Interaction of Innovation with Legal Origins}
    \label{table:fpi_legal}
    \end{threeparttable}}
\end{table}

The lack of statistical significance in the German legal origin and Socialist legal origin groups for debt inflow should be interpreted with caution and is likely influenced by classification and sample composition considerations, rather than an absence of an innovation–FPI linkage. An important empirical issue is that China is classified under the German legal origin in standard legal-origin datasets.\footnote{For details on classification of legal origins, please see \cite{la2017legal}.} Because China is an outlier in scale and in the structure of capital account controls and market access over parts of the sample, its inclusion in the German origin group can significantly affect estimates and precision. In particular, foreign portfolio inflows in China are shaped by regulations, liberalization, and distinct institutional arrangements that may weaken the contemporaneous mapping from patenting to foreign portfolio inflows in the country-year data. This composition effect can reduce the estimated sensitivity or increase uncertainty for the whole estimation group.

Furthermore, the Socialist legal origin category in the sample is extremely small and concentrated. In classifications, only a limited number of countries fall into this group; in the present dataset, it is only driven by two cases (North Korea and Myanmar). With such limited within-group variation, the interaction estimates for Socialist legal origin inevitably have low statistical power. Therefore, an insignificant coefficient for this group is more plausibly interpreted as an imprecision problem than as evidence that innovation does not matter for FPI.

\section{Local Projection with the Instrumental Variable}\label{sec:dynamic_fpi}

While the previous analysis establishes the mean and nonlinear effect of innovation on FPI, it remains unclear how quickly or persistently these effects unfold over time. Debt and equity inflows may respond to innovation with delays due to legal protections, contract enforcement, adjustment costs, institutional frictions, or the time required for new technologies to diffuse. Static models capture only the contemporaneous average effect, potentially ignoring important dynamic paths. In this section, we implement the LP framework following \cite{jorda2005estimation}, extending it with the same instruments to address endogeneity concerns. Unlike vector autoregressions (VARs), the LP method does not require imposing strong assumptions on the joint dynamics of the system and is suitable for estimating impulse response functions in panel data with heterogeneous units. We estimate the specification for each horizon $h = 0, 1, \ldots, H$ given by
\begin{equation}
	\ln (y^s_{i,t+h}) = \rho^{h} \ln (x_{i,t}) + \varphi^{h} \mathbf{W}_{i,t} + \mu_t^{h} +  \pi_r^{h} + \varepsilon_{i,t+h}^{h}
\end{equation}
where $y^s_{i,t+h}$ denotes either debt inflow, equity inflow, total FPI, or the E/D inflow ratio in the country $i$ at time $t+h$; $x_{i,t}$ is the instrumented value of innovation intensity at time $t$; $\mathbf{W}_{it}$ represents a vector of control variables; time fixed effects $\mu_t^{h}$ control for the common shocks; regional fixed effects are captured by $\pi_r^{h}$. The coefficient of interest, $\rho^{h}$, captures the period-specific response of FPI for $h$ periods after an innovation shock at time $t$. We estimate using White’s heteroskedasticity-robust standard error \citep{white1980heteroskedasticity}.

\begin{figure}[ht]
	\begin{center}
		\includegraphics[width=1.0\textwidth]{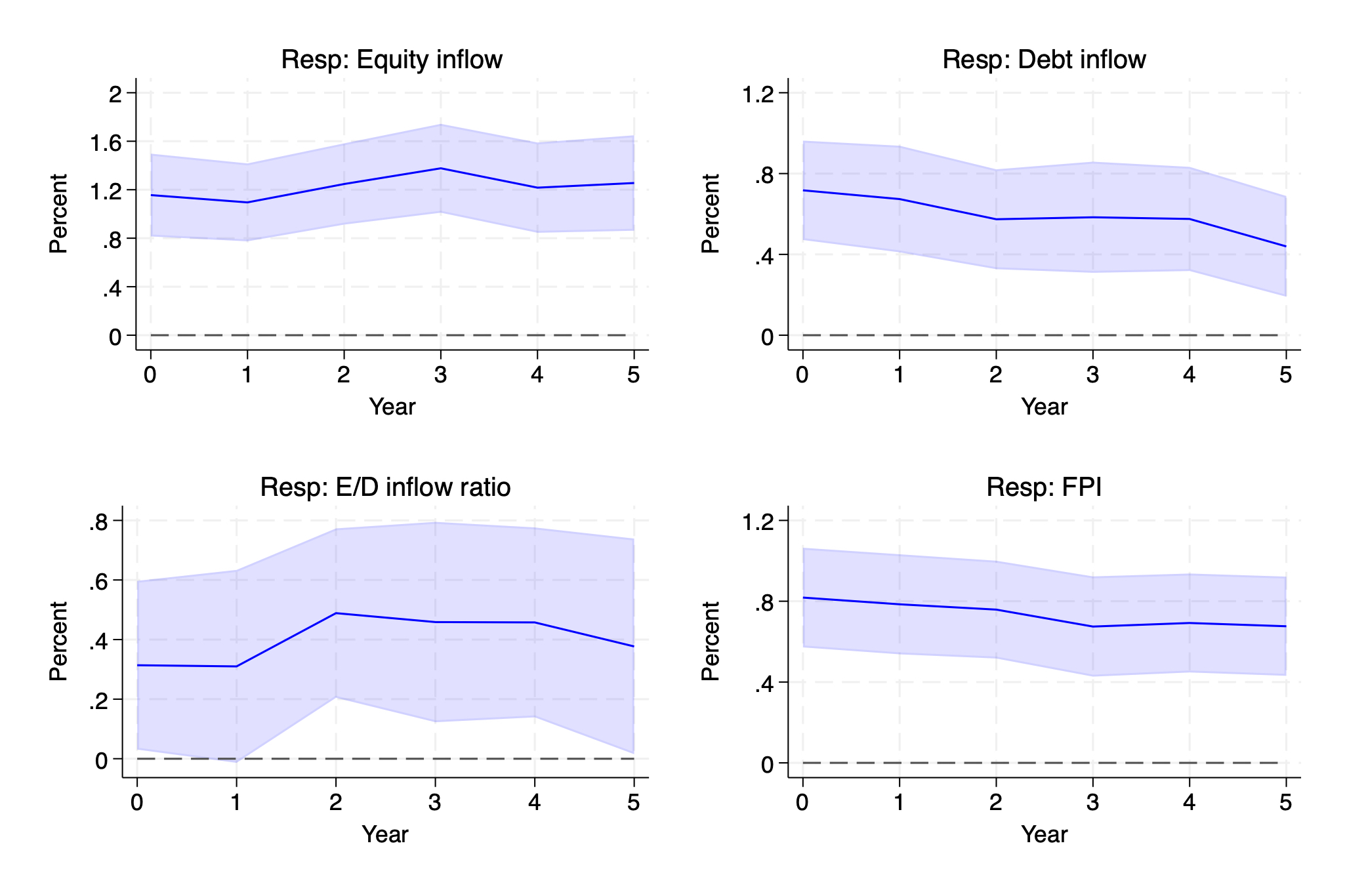}
		\caption{Impulse Response to Innovation }
		\label{figure:lpiv_fpi}
	\end{center}
	\floatfoot{\footnotesize\textit{Note}: Impulse-response functions (IRF) showing the response of debt inflow (right-top), equity inflow (left-top), E/D inflow ratio (left-bottom), and total FPI (right-bottom). The blue solid line represents the response of the dependent variable to an increase in patents for the forecast horizon $h=0, 1, 2, 3, 4,...,5$. Blue shadow band represents the 95\% confidence interval calculated based on White's heteroskedasticity-robust standard error. The horizontal axis represents the year after an increase in patent applications. Instruments and control variables are the same as the baseline IV specification and include time fixed effects.}
\end{figure}

The LP estimates in Figure \ref{figure:lpiv_fpi} trace how a rise in innovation intensity impacts portfolio inflows over the next five years. The response of equity inflow is immediate and persistent: the point effect rises from roughly 1.1 percent on impact to around 1.3 percent by year 3–5, and the confidence bands remain above zero at all horizons. By contrast, debt inflow reacts in a monotonically decreasing way. The point effect persistently declines from roughly 0.7 percent in the initial year on impact to around 0.4 percent in the final year. Composition effects are explicit in the E/D inflow ratio panel. The ratio rises steadily from the initial year, peaks around year 2 (0.5 percent), and remains above zero at horizon 5, indicating a durable shift of inflows toward equity following an innovation shock.

The combined response of total FPI mirrors the debt inflow pattern, but with a larger magnitude: an initial jump near 0.9 percent that then drifts down over subsequent years. Together, these paths suggest that innovation is quickly capitalized by foreign equity investors, who can reprice and scale exposures, while debt inflow responds more cautiously and temperately, likely reflecting contract rigidity and rollover considerations. All responses are estimated conditional on controls and horizon-specific region and time fixed effects, so the dynamics are not driven by common global shocks.

\section{Conclusion} \label{sec:conclusion_fpi}

Our baseline findings show that innovation is a powerful predictor of FPI. Instruments based on regional innovation spillovers and a global research position push are strongly relevant and pass weak-IV diagnostics, allowing a causal interpretation of second-stage coefficients. Moreover, we find that the rise in patent applications per million people is associated with a substantially larger rise in equity inflow than in debt inflow, implying a composition asymmetry: innovation attracts risky portfolio capital more significantly through equity. These findings are consistent across specifications with and without the full control set, as well as across alternative definitions of the independent and dependent variables.  

Beyond average effects, the innovation-FPI link is state-dependent. When levels of technological development and institutional quality are low to moderate, innovation mainly tilts FPI toward equity; only after countries approach the top tercile of institutional quality does debt inflow respond significantly. This sequencing is consistent with the argument that equity investors value growth options and can reallocate more flexibly in response to new information, whereas sustained debt inflow requires very high levels of democracy, contract and law enforcement, corruption control, and administrative capacity for long-term, stable investments. The grouped estimates across technological development and institutional indicators reveal a nearly monotonic equity inflow gradient with improvements but a threshold pattern for debt inflow. These patterns help explain the Lucas paradox in FPI: innovation-intensive, well-governed economies capture the largest share of global portfolio capital, in both debt and equity inflows.

Finally, local projection with instrumental variables indicates that equity inflow reacts immediately and persistently to innovation shocks over a multi-year horizon, whereas the debt inflow response is smaller and dampens over time. The total FPI path mirrors debt inflow, and the E/D inflow ratio remains positive durably, signaling a persistent compositional shift following innovation impulses. These results suggest that international investors quickly capitalize on new information about innovation intensity by scaling equity exposures, while debt inflow adjusts more cautiously depending on the aggregate conditions of the target countries.

One limitation of this study stems from the nature of the FPI data, which do not allow a detailed decomposition of investor types or investment motivations. While the CPIS measures provide comprehensive and internationally comparable information on equity and debt inflows, they do not distinguish between institutional and individual investors, nor do they separate long-term strategic holdings from short-term portfolio reallocations within the same asset class. As a result, the analysis captures the aggregate response of foreign portfolio inflows to domestic innovation, rather than the potentially heterogeneous reactions of different investor groups.

\clearpage
\newpage
\appendix
\renewcommand{\thetable}{A\arabic{table}}
\setcounter{table}{0}

\section{Robustness Checks}\label{sec:robustness_fpi}

This section presents several primary robustness checks, demonstrating that the estimated relationship between innovation and FPI remains stable across key alternative specifications. 

\subsection{Alternative Innovation Proxies}\label{ipcharges}

\begin{table}[ht]
	\centering
	\footnotesize
	\makebox[\textwidth][c]{%
		\begin{threeparttable}
			\begin{tabular}{
					p{2.5cm} 
					>{\centering\arraybackslash}m{1cm} 
					>{\centering\arraybackslash}m{1cm}
					>{\centering\arraybackslash}m{1cm}
					>{\centering\arraybackslash}m{1cm}
					>{\centering\arraybackslash}m{1cm}
					>{\centering\arraybackslash}m{1cm}
					>{\centering\arraybackslash}m{1cm}
					>{\centering\arraybackslash}m{1cm}
					>{\centering\arraybackslash}m{1cm}
				} 
				\toprule
				& \multicolumn{3}{c}{IP Receipts} 
				& \multicolumn{3}{c}{High-tech exports} 
				& \multicolumn{3}{c}{Total patents} \\
				\cmidrule(lr){2-4}
				\cmidrule(lr){5-7}
				\cmidrule(lr){8-10}
				& (1a) & (1b) & (1c) 
				& (2a) & (2b) & (2c) 
				& (3a) & (3b) & (3c) \\
				\midrule
				\underline{2nd Stage} & & & & & & & & & \\
				Equity inflow & 
				0.601*** &        &        &
				1.068*** &        &        &
				1.592*** &        &        \\
				& 
				(0.092) &        &        &
				(0.225) &        &        &
				(0.258) &        &        \\
				Debt inflow & 
				& 0.358*** &        &
				& 0.571*** &        &
				& 0.958*** &        \\
				& 
				& (0.073) &        &
				& (0.189) &        &
				& (0.173) &        \\
				E/D inflow ratio & 
				&        & 0.216** &
				&        & 0.460** &
				&        & 0.423** \\
				& 
				&        & (0.094) &
				&        & (0.181) &
				&        & (0.191) \\

				\underline{Weak IV Test} & & & & & & & & & \\
				CLR & 
				0.000 & 0.000 & 0.023 
				& 0.000 & 0.000 & 0.013 
				& 0.000 & 0.000 & 0.028 \\

				AR & 
				0.000 & 0.000 & 0.061 
				& 0.000 & 0.001 & 0.041 
				& 0.000 & 0.000 & 0.086 \\

				Wald & 
				0.000 & 0.000 & 0.021 
				& 0.000 & 0.003 & 0.011 
				& 0.000 & 0.000 & 0.027 \\
				\midrule
				Observations & 
				438 & 443 & 384 
				& 274 & 279 & 253 
				& 500 & 499 & 438 \\

				Time period & 
				1996--2021 & 1996--2021 & 1996--2021 
				& 2007--2021 & 2007--2021 & 2007--2021 
				& 1996--2021 & 1996--2021 & 1996--2021 \\

				Time FE & 
				YES & YES & YES 
				& YES & YES & YES 
				& YES & YES & YES \\
				Region FE & 
				YES & YES & YES 
				& YES & YES & YES 
				& YES & YES & YES \\
				Controls & 
				YES & YES & YES 
				& YES & YES & YES 
				& YES & YES & YES \\
				R-squared & 
				0.387 & 0.131 & 0.290 
				& 0.002 & 0.058 & 0.201 
				& 0.086 & 0.052 & 0.212 \\
				[1ex] 
				\bottomrule
			\end{tabular}
			\begin{tablenotes}
				\footnotesize
				\item The dependent variable is either equity inflow, debt inflow, or the E/D inflow ratio. The endogenous variable is the alternative innovation proxy. Columns (1a)--(1c) show receipts for intellectual property rights, columns (2a)--(2c) show high-tech exports, and columns (3a)--(3c) show total patents. Statistical significance: *** \( p < 0.01 \), ** \( p < 0.05 \), * \( p < 0.1 \). Robust standard errors in parentheses. AR and Wald tests follow \cite{olea2013robust}. See \cite{pflueger2015robust} for discussions of weak instrument tests and \cite{finlay2014weakiv10} for Stata implementations. \(P\)-values are reported for CLR, AR, and Wald tests of weak instruments.
			\end{tablenotes}
			\caption{Alternative Measures of Innovation}
			\label{table:innov_fpi}
	\end{threeparttable}}
\end{table}

To avoid relying solely on patent applications as the measure of innovation, we further incorporate innovation proxies at the country level in Table \ref{table:innov_fpi}. Specifically, we employ three alternative indicators: (1) total charges (in billion USD) per million people for the authorized use of intellectual property (IP) rights, including patents, trademarks, copyrights, trade secrets, and industrial processes; (2) high-technology exports (in billion USD) per million people, encompassing products with high R\&D intensity, such as aerospace, computers, pharmaceuticals, and scientific instruments; and (3) an extended measure of patent applications per million people that includes both resident and non-resident filings. Our findings remain robust and consistent with the baseline results using the same identification strategy, which supports the role of innovation in shaping FPI. 
Table \ref{table:innov_fpi} shows that alternative innovation measures all exhibit positive and significant effects on FPI, though magnitudes vary. In general, comprehensive innovation indicators suggest that innovation robustly promotes portfolio allocation by international investors.

\subsection{Alternative Instruments}

We employ alternative approaches to measure the geographic distance between countries and, to check our primary findings, construct three additional instruments based on the following methods. Three additional instruments differ in their distance measures. The first is to use the distance between the two capitals of the countries instead of the distances between the largest cities. The other two measures are weighted calculations. To quantify the weighted distances between countries, we follow the methodology of \cite{head2002illusory}. Let $d_{k,\ell}$ denote the distance between agglomeration $k$ in country $i$ and agglomeration $\ell$ in country $j$. The population-weighted average distance between countries $i$ and $j$ is computed as
\begin{equation*}
	d_{i,j} = \left( \sum_{k \in i} \frac{\text{pop}_k}{\text{pop}_i} \sum_{\ell \in j} \frac{\text{pop}_\ell}{\text{pop}_j} \, d_{k,\ell}^{\theta} \right)^{1/\theta}
\end{equation*}
where $\text{pop}_k$ and $\text{pop}_\ell$ denote the populations of agglomerations $k$ and $\ell$, respectively, and $\theta$ determines how sensitive trade flows are to distance.

For the weighted measure in terms of population, we set $\theta = 1$, producing a simple arithmetic mean of bilateral distances weighted by population shares. By contrast, for the last weighted distance measure, we set $\theta = -1$, which produces a harmonic mean of bilateral distances. The harmonic mean assigns greater weight to shorter distances and aligns with the negative coefficients typically observed for distance in gravity models of bilateral trade. Consequently, using $\theta = -1$ better captures the economic importance of proximity, reflecting the fact that nearby regions tend to exhibit disproportionately strong trade and economic linkages.

Table \ref{table:instruments_fpi} presents the results from using alternative instruments for the number of patent applications per million people, examining whether the baseline findings are robust to different definitions of geographic distance. Across all specifications, the estimated coefficients on FPI remain positive and statistically significant, confirming the baseline result that innovation is positively associated with FPI.

\begin{table}[ht]
	\centering
	\footnotesize
	\makebox[\textwidth][c]{%
		\begin{threeparttable}
			\begin{tabular}{
					p{2.5cm} 
					>{\centering\arraybackslash}m{1cm} 
					>{\centering\arraybackslash}m{1cm}
					>{\centering\arraybackslash}m{1cm}
					>{\centering\arraybackslash}m{1cm}
					>{\centering\arraybackslash}m{1cm}
					>{\centering\arraybackslash}m{1cm}
					>{\centering\arraybackslash}m{1cm}
					>{\centering\arraybackslash}m{1cm}
					>{\centering\arraybackslash}m{1cm}
				} 
				\toprule
				& \multicolumn{3}{c}{Capital distance IV} 
				& \multicolumn{3}{c}{Weighted distance IV} 
				& \multicolumn{3}{c}{Economic proximity IV} \\
				\cmidrule(lr){2-4} \cmidrule(lr){5-7} \cmidrule(lr){8-10}
				& (1a) & (1b) & (1c) & (2a) & (2b) & (2c) & (3a) & (3b) & (3c) \\
				\midrule
				\underline{2nd Stage} & & & & & & & & & \\
				Equity inflow & 
				1.156***  &        &        &
				1.153***   &        &        &
				1.152***   &        &        \\
				& 
				(0.174)   &        &        &
				(0.174)   &        &        &
				(0.174)   &        &        \\
				Debt inflow & 
				& 0.717*** &        &
				& 0.716***  &        &
				& 0.715***  &        \\
				& 
				& (0.125)  &        &
				& (0.125)  &        &
				& (0.125)  &        \\
				E/D inflow ratio & 
				&        & 0.314**  &
				&        & 0.313**  &
				&        & 0.313**  \\
				& 
				&        & (0.145)   &
				&        & (0.144)   &
				&        & (0.144)   \\
				\underline{Weak IV Test} & & & & & & & & & \\
				CLR  & 
				0.000 & 0.000 & 0.031 &
				0.000 & 0.000 & 0.031 &
				0.000 & 0.000 & 0.031 \\
				AR   & 
				0.000 & 0.000 & 0.096 &
				0.000 & 0.000 & 0.096 &
				0.000 & 0.000 & 0.096 \\
				Wald & 
				0.000 & 0.000 & 0.030 &
				0.000 & 0.000 & 0.030 &
				0.000 & 0.000 & 0.030 \\
				\midrule
				Observations & 
				501 & 500 & 439 &
				501 & 500 & 439 &
				501 & 500 & 439 \\
				Time period & 
				1996--2021 & 1996--2021 & 1996--2021 &
				1996--2021 & 1996--2021 & 1996--2021 & 
				1996--2021 & 1996--2021 & 1996--2021 \\
				Time FE & 
				YES & YES & YES & 
				YES & YES & YES & 
				YES & YES & YES \\
				Region FE & 
				YES & YES & YES & 
				YES & YES & YES & 
				YES & YES & YES \\
				Controls & 
				YES & YES & YES & 
				YES & YES & YES & 
				YES & YES & YES \\
				R-squared & 
				0.191 & 0.037 & 0.231 & 
				0.192 & 0.038 & 0.231 &
				0.193 & 0.038 & 0.231 \\
				[1ex] 
				\bottomrule
			\end{tabular}
			\begin{tablenotes}
				\footnotesize
				\item The dependent variable is either equity inflow, debt inflow, or the E/D inflow ratio. The endogenous variable is the number of patent applications instrumented with the alternative instruments and regional dummies. Columns (1a)--(1c) use the capital distance IV. Columns (2a)--(2c) use the weighted distance IV. Columns (3a)--(3c) use the economic proximity IV. Statistical significance: *** $p<0.01$, ** $p<0.05$, * $p<0.1$. Robust standard errors in parentheses. AR and Wald tests follow \cite{olea2013robust}. See \cite{pflueger2015robust} for discussions of weak instrument tests and \cite{finlay2014weakiv10} for Stata implementations. $P$-values are reported for CLR, AR, and Wald tests of weak instruments.
			\end{tablenotes}
			\caption{Alternative Regional Push Instruments}
			\label{table:instruments_fpi}
	\end{threeparttable}}
\end{table}

Specifically, for debt inflow, the coefficients remain smaller than equity inflow, consistently around 0.72 versus 1.15 with significance at the 1\% level, suggesting that the relationship between innovation and FPI is stable regardless of the distance measure used. For the E/D inflow ratio, we also observe significant coefficients in all three instruments. Weak instrument diagnostics (CLR, AR, and Wald tests) consistently produce highly significant p-values close to zero for most specifications, indicating that the instruments are not weak and provide credibility to the IV estimates. 

\subsection{Placebo Test}
Finally, we present a placebo test by regressing our innovation intensity on past levels of three indicators of FPI to address the concern of unobserved common causal factors. As presented in Table \ref{table:placebo_fpi}, the coefficients on the first and second lag of the three indicators of FPI are insignificant, which confirms our main results on the causal relationship from innovation intensity to FPI.
\begin{table}[ht]
	\centering
	\footnotesize
	\begin{threeparttable}
		\begin{tabular}{
				>{\raggedright\arraybackslash}p{4.5cm}
				*{3}{>{\centering\arraybackslash}p{1cm}}
			}
			\toprule
			& \multicolumn{3}{c}{Innovation Intensity} \\
			\cmidrule(lr){2-4}
			& (1) & (2) & (3) \\
			\midrule
			Lagged dependent variable & 0.886*** (0.105) & 0.674*** (0.216) & 0.851*** (0.118) \\
			Equity inflow $(t-1)$ & 0.009 \quad(0.061) & & \\
			Equity inflow $(t-2)$ & 0.002 \quad(0.066) & & \\
			Debt inflow $(t-1)$ & & -0.059\quad(0.051) & \\
			Debt inflow $(t-2)$ & & 0.122 \quad(0.076) & \\
			E/D inflow ratio $(t-1)$ & & & 0.025 \quad(0.044) \\
			E/D inflow ratio $(t-2)$ & & & -0.028 \quad(0.050) \\
			\midrule
			Observations & 726 & 747 & 514 \\
			Number of countries & 65 & 60 & 61 \\
			Time FE & YES & YES & YES \\
			Rgion FE & YES & YES & YES \\
			[1ex]
			\bottomrule
		\end{tabular}
		\begin{tablenotes}
			\footnotesize
			\item Note: The dependent variable is the innovation intensity. Statistical significance levels are indicated by the asterisks: *** \( p < 0.01 \), ** \( p < 0.05 \), * \( p < 0.1 \). All estimations are based on the system GMM estimator. Robust standard errors are in parentheses.
		\end{tablenotes}
		\caption{Placebo Test}
		\label{table:placebo_fpi}
	\end{threeparttable}
\end{table}



\clearpage

\section{Alternative samples and methods}

In addition to the three primary robustness checks for the baseline model, we perform four further analyses to confirm the robustness of the relationship between patent applications and FPI. Table \ref{table:regional_leaders_fpi} reports results by instruments reconstructed using time-varying regional innovation leaders, which select the maximum number of patent application holders in each year within the region. We run the specification without controlling for years of banking crisis in Table \ref{table:bankingcrisis}. 
Table \ref{table:stderror} reports estimation results under alternative standard error assumptions, comparing conventional and clustered by year standard errors. Table \ref{table:winsorized} presents results using winsorized measures of FPI to reduce the influence of outliers. Across all specifications and subsamples, the estimated effects of innovation on FPI remain positive and statistically significant, enhancing the robustness of the baseline findings.

\begin{table}[ht]
	\centering
	\footnotesize
	\makebox[\textwidth][c]{%
		\begin{threeparttable}
			\begin{tabular}{
					p{2.5cm} 
					>{\centering\arraybackslash}m{0.8cm} 
					>{\centering\arraybackslash}m{0.8cm}
					>{\centering\arraybackslash}m{0.8cm}
					>{\centering\arraybackslash}m{0.8cm}
					>{\centering\arraybackslash}m{0.8cm}
					>{\centering\arraybackslash}m{0.8cm}
					>{\centering\arraybackslash}m{0.8cm}
					>{\centering\arraybackslash}m{0.8cm}
					>{\centering\arraybackslash}m{0.8cm}
					>{\centering\arraybackslash}m{0.8cm}
					>{\centering\arraybackslash}m{0.8cm}
					>{\centering\arraybackslash}m{0.8cm}
				} 
				\toprule
				& \multicolumn{3}{c}{Baseline IV} 
				& \multicolumn{3}{c}{Capital IV} 
				& \multicolumn{3}{c}{Weighted IV}
				& \multicolumn{3}{c}{Economic IV} \\
				\cmidrule(lr){2-4} 
				\cmidrule(lr){5-7} 
				\cmidrule(lr){8-10}
				\cmidrule(lr){11-13}
				& (1a) & (1b) & (1c) 
				& (2a) & (2b) & (2c) 
				& (3a) & (3b) & (3c)
				& (4a) & (4b) & (4c) \\
				\midrule
				\underline{2nd Stage} & & & & & & & & & & & & \\
				Equity inflow & 
				1.095*** &        &        &
				1.095*** &        &        &
				1.094*** &        &        &
				1.093*** &        &        \\
				& 
				(0.170) &        &        &
				(0.170) &        &        &
				(0.170) &        &        &
				(0.170) &        &        \\
				Debt inflow & 
				& 0.713*** &        &
				& 0.713*** &        &
				& 0.714*** &        &
				& 0.714*** &        \\
				& 
				& (0.123) &        &
				& (0.123) &        &
				& (0.123) &        &
				& (0.123) &        \\
				E/D inflow ratio & 
				&        & 0.310** &
				&        & 0.310** &
				&        & 0.309** &
				&        & 0.308** \\
				& 
				&        & (0.146) &
				&        & (0.146) &
				&        & (0.146) &
				&        & (0.146) \\

				\underline{Weak IV Test} & & & & & & & & & & & & \\
				CLR & 
				0.000 & 0.000 & 0.041 &
				0.000 & 0.000 & 0.041 &
				0.000 & 0.000 & 0.042 &
				0.000 & 0.000 & 0.043 \\

				AR & 
				0.000 & 0.000 & 0.091 &
				0.000 & 0.000 & 0.091 &
				0.000 & 0.000 & 0.090 &
				0.000 & 0.000 & 0.089 \\

				Wald & 
				0.000 & 0.000 & 0.033 &
				0.000 & 0.000 & 0.033 &
				0.000 & 0.000 & 0.034 &
				0.000 & 0.000 & 0.034 \\
				\midrule
				Observations & 
				513 & 513 & 452 &
				513 & 513 & 452 &
				513 & 513 & 452 &
				513 & 513 & 452 \\

				Time period & 
				1996--2021 & 1996--2021 & 1996--2021 &
				1996--2021 & 1996--2021 & 1996--2021 &
				1996--2021 & 1996--2021 & 1996--2021 &
				1996--2021 & 1996--2021 & 1996--2021 \\

				Time FE & 
				YES & YES & YES &
				YES & YES & YES &
				YES & YES & YES &
				YES & YES & YES \\
				Region FE & 
				YES & YES & YES &
				YES & YES & YES &
				YES & YES & YES &
				YES & YES & YES \\
				Controls & 
				YES & YES & YES &
				YES & YES & YES &
				YES & YES & YES &
				YES & YES & YES \\
				R-squared & 
				0.235 & 0.044 & 0.234 &
				0.235 & 0.044 & 0.234 &
				0.235 & 0.044 & 0.234 &
				0.235 & 0.044 & 0.234 \\
				[1ex] 
				\bottomrule
			\end{tabular}
			\begin{tablenotes}
				\footnotesize
				\item The dependent variable is either equity inflow, debt inflow, or the E/D inflow ratio. The endogenous variable is the number of patent applications instrumented with alternative regional leader instruments. Columns (1a) to (1c) use the baseline IV. Columns (2a) to (2c) use the capital distance IV. Columns (3a) to (3c) use the weighted distance IV. Columns (4a) to (4c) use the economic proximity IV Statistical significance: *** $p<0.01$, ** $p<0.05$, * $p<0.1$. Robust standard errors in parentheses. AR and Wald tests follow \cite{olea2013robust}. See \cite{pflueger2015robust} for discussions of weak instrument tests and \cite{finlay2014weakiv10} for Stata implementations. $P$-values are reported for CLR, AR, and Wald tests of weak instruments.
			\end{tablenotes}
			\caption{Time-varying Regional Leaders}
			\label{table:regional_leaders_fpi}
	\end{threeparttable}}
\end{table}

\begin{table}[ht]
	\centering
	\footnotesize
	\begin{threeparttable}
		\begin{tabular}{
				>{\raggedright\arraybackslash}p{4cm} 
				>{\centering\arraybackslash}m{2.6cm} 
				>{\centering\arraybackslash}m{2.7cm}
				>{\centering\arraybackslash}m{2.6cm}
			} 
			\toprule
			& Equity inflow & Debt inflow &  E/D inflow ratio \\
			\cmidrule(lr){2-2} \cmidrule(lr){3-3} \cmidrule(lr){4-4}
			& (1) & (2) & (3) \\
			\midrule

			\underline{2nd Stage} &  &  &  \\

			Patents & 
			1.101***  & 0.641***  & 0.309** \\
        & (0.176) &  (0.117) &  (0.147) 			\\
			\underline{1st Stage} &  &  &  \\
        Regional push instrument& 
        -0.123***  & -0.122*** & -0.161***  \\
		& 
		 (0.039) & (0.036) &  (0.038) \\
        Global push instrument & 
        3.204***  & 3.339*** & 3.293***  \\
		& 
		 (0.229) & (0.209) &  (0.217) \\

			\underline{Weak IV Test} &  &  &  \\ 

			CLR  & 
			0.000 & 0.000 & 0.034 \\

			AR  & 
			0.000 & 0.000 & 0.104 \\

			Wald  & 
			0.000 & 0.000 & 0.036 \\

			\midrule
			Period & 1996-2021 & 1996-2021 & 1996-2021 \\
			Time FE & YES & YES & YES \\
            Region FE & YES & YES & YES \\
			Controls & YES & YES & YES \\
			Obs. & 529 & 538 & 467 \\
			R-squared & 0.196 & 0.046 & 0.230 \\
			[1ex] 
			\bottomrule
		\end{tabular}
		\begin{tablenotes}
			\footnotesize
			\item The dependent variable is either equity inflow, debt inflow, or the E/D inflow ratio. The endogenous variable is the number of patent applications instrumented with the regional innovation spillover instrument and regional dummies. Control variables are in Table \ref{table:data_description}. Note: *** \( p < 0.01 \), ** \( p < 0.05 \), * \( p < 0.1 \). Numbers in parentheses are robust standard errors. AR and Wald tests follow the procedures in \cite{olea2013robust}. Multiple IVs yield extra CLR statistics; see \cite{pflueger2015robust} for discussions of weak instrument tests in linear IV regressions and \cite{finlay2014weakiv10} for Stata implementations. $P$-values are reported for CLR, AR, and Wald tests.
		\end{tablenotes}
		\caption{Dropping banking crisis years}
		\label{table:bankingcrisis}
	\end{threeparttable}
\end{table}

\clearpage

\begin{table}[ht]
    \centering
    \footnotesize
    \begin{threeparttable}
    \begin{tabular}{
        >{\raggedright\arraybackslash}p{2.3cm} 
        >{\centering\arraybackslash}m{1.5cm} 
        >{\centering\arraybackslash}m{1.5cm}
        >{\centering\arraybackslash}m{1.5cm} 
        >{\centering\arraybackslash}m{1.5cm} 
        >{\centering\arraybackslash}m{1.5cm}
        >{\centering\arraybackslash}m{1.5cm}
    } 
         \toprule
         & \multicolumn{2}{c}{Equity inflow} & \multicolumn{2}{c}{Debt inflow} & 
         \multicolumn{2}{c}{E/D inflow ratio} \\
         \cmidrule(lr){2-3} \cmidrule(lr){4-5} \cmidrule(lr){6-7}
         & (1) & (2) & (3) & (4) & (5) & (6) \\
         \midrule

        \underline{2nd Stage} &  &  &  &  &  & \\

        Patents & 
        1.156***  & 1.156***  & 
        0.717***  & 0.717***  & 
        0.314** & 0.314***  \\

         & 
         (0.153) & (0.121) & 
        (0.143) &  (0.097) & 
         (0.157) &  (0.112) \\

        \underline{Weak IV Test} &  &  &  &  &  & \\ 

        CLR  & 
        0.000 & 0.000 & 
        0.000 & 0.000 & 
        0.048 & 0.034 \\

        AR  & 
        0.000 & 0.000 & 
        0.000 & 0.000 & 
        0.142 & 0.090 \\

        Wald  & 
        0.000 & 0.000 & 
        0.000 & 0.000 & 
        0.045 & 0.005 \\

        \midrule
        Period & 1996-2021 & 1996-2021 & 1996-2021 & 1996-2021 & 1996-2021 & 1996-2021 \\
        Time FE & YES & YES & YES & YES & YES & YES \\
        Region FE & YES & YES & YES & YES & YES & YES \\
        Controls & YES & YES & YES & YES & YES & YES \\
        Obs. & 501 & 501 & 500 & 500 & 439 & 439 \\
        R-squared & 0.191 & 0.191 & 0.037 & 0.037 & 0.231 & 0.231 \\
       [1ex] 
       \bottomrule
    \end{tabular}
    \begin{tablenotes}
         \footnotesize
        \item The dependent variable is either equity inflow, debt inflow, or the E/D inflow ratio. The endogenous variable is the number of patent applications per million people. Control variables are in Table \ref{table:data_description}. Note: *** \( p < 0.01 \), ** \( p < 0.05 \), * \( p < 0.1 \). Columns (1), (3), and (5) are estimates using normal standard errors. Columns (2), (4), and (6) are estimates using standard errors clustered by year. AR and Wald tests follow the procedures in \cite{olea2013robust}. Multiple IVs yield extra CLR statistics; see \cite{pflueger2015robust} for discussions of weak instrument tests in linear IV regressions and \cite{finlay2014weakiv10} for Stata implementations. $P$-values are reported for CLR, AR, and Wald tests.
    \end{tablenotes}
    \caption{Baseline model with normal and clustered standard errors}
    \label{table:stderror}
\end{threeparttable}
\end{table}

\begin{table}[ht]
    \centering
    \footnotesize
    \begin{threeparttable}
    \begin{tabular}{
        >{\raggedright\arraybackslash}p{4cm} 
        >{\centering\arraybackslash}m{2.5cm} 
        >{\centering\arraybackslash}m{2.5cm}
        >{\centering\arraybackslash}m{2.6cm}
    } 
         \toprule
         & Equity inflow & Debt inflow & E/D inflow ratio\\
         \cmidrule(lr){2-2} \cmidrule(lr){3-3} \cmidrule(lr){4-4}
         & (1) & (2) & (3) \\
         \midrule

        \underline{2nd Stage} &  &  &  \\

        Patents & 
        1.052***  & 0.633***  & 0.333**  \\
        & 
         (0.158) &  (0.111) &  (0.140) \\

        \underline{1st Stage} &  &  &  \\
        Regional push instrument& 
        -0.153***  & -0.170*** & -0.179***  \\
		& 
		 (0.044) & (0.037) &  (0.040) \\
        Global push instrument & 
        3.287***  & 3.462*** & 3.448***  \\
		& 
		 (0.236) & (0.215) &  (0.223) \\

        \underline{Weak IV Test} &  &  &  \\ 

        CLR  & 
        0.000 & 0.000 & 0.019 \\

        AR  & 
        0.000 & 0.000 & 0.061 \\

        Wald  & 
        0.000 & 0.000 & 0.018 \\

        \midrule
        Period & 1996-2021 & 1996-2021 & 1996-2021 \\
        Time FE & YES & YES & YES \\
        Controls & YES & YES & YES \\
        Obs. & 501 & 500 & 439 \\
        R-squared & 0.199 & 0.040 & 0.245 \\
       [1ex] 
       \bottomrule
    \end{tabular}
    \begin{tablenotes}
        \footnotesize
        \item The dependent variable is either equity inflow, debt inflow, or the E/D inflow ratio, using winsorized baseline variables at the 1\% level. The endogenous variable is the winsorized number of patent applications at 1\% level. Control variables are in Table \ref{table:data_description}. Note: *** \( p < 0.01 \), ** \( p < 0.05 \), * \( p < 0.1 \). Numbers in parentheses are robust standard errors. AR and Wald tests follow the procedures in \cite{olea2013robust}. Multiple IVs yield extra CLR statistics; see \cite{pflueger2015robust} for discussions of weak instrument tests in linear IV regressions and \cite{finlay2014weakiv10} for Stata implementations. $P$-values are reported for CLR, AR, and Wald tests.
    \end{tablenotes}
    \caption{Effect of innovation on FPI: Winsorized sample}
    \label{table:winsorized}
\end{threeparttable}
\end{table}

\clearpage
\newpage

\bibliographystyle{apalike} 
\bibliography{ref}

\clearpage

\end{document}